\documentclass[%
 reprint,
superscriptaddress,
 amsmath,amssymb,
 aps,
]{revtex4-2}
\pdfoutput=1
\usepackage{graphicx}
\usepackage{dcolumn}
\usepackage{bm}
\usepackage[table]{xcolor}
\usepackage{colortbl}

\usepackage[labelformat=simple, caption=false,font={bf}]{subfig}
\usepackage{color}
\graphicspath{ {pictures/} }

\renewcommand{\thetable}{\arabic{table}}

\begin{document}


\title{Predicting attractors from spectral properties of stylized gene regulatory networks}

\author{Dzmitry Rumiantsau}
 \email{d.rumiantsau@constructor.university}
\affiliation{%
Department of Life Sciences and Chemistry, Constructor University, D-28759 Bremen, Germany}
\author{ Annick Lesne}%
 \email{annick.lesne@sorbonne-universite.fr}
\affiliation{%
 Sorbonne Universit\'e, CNRS,  Laboratoire de Physique Th\'eorique de la Mati\`ere Condens\'ee, LPTMC, F-75252, Paris, France}%
\affiliation{%
	Institut de G\'en\'etique Mol\'eculaire de Montpellier, University of Montpellier, CNRS, F-34293, Montpellier, France\looseness=-1}%

\author{Marc-Thorsten H\"utt}
 \email{m.huett@constructor.university}
\affiliation{Department of Life Sciences and Chemistry, Constructor University, D-28759 Bremen, Germany}

\date{\today}

\begin{abstract}
How the architecture of gene regulatory networks ultimately shapes gene expression patterns is an open question, which has been approached from a multitude of angles. The dominant strategy has been to identify non-random features in these networks and then argue for the function of these features using mechanistic modelling. Here we establish the foundation of an alternative approach by studying the correlation of eigenvectors with synthetic gene expression data simulated with a basic and popular model of gene expression dynamics -- attractors of Boolean threshold dynamics in signed directed graphs.

Eigenvectors of the graph Laplacian are known to explain collective dynamical states (stationary patterns) in Turing dynamics on graphs. In this study, we show that eigenvectors can also predict collective states  (attractors) for a markedly different type of dynamics, Boolean threshold dynamics, and category of graphs, signed directed graphs. However, the overall predictive power depends on details of the network architecture, in a predictable fashion.

Our results are a set of statistical observations, providing the first systematic step towards a further theoretical understanding of the role of eigenvectors in dynamics on graphs. 

\end{abstract}

\maketitle


\noindent

\section{\label{sec:intro}Introduction}

Understanding how collective dynamical states can be predicted just from network architecture remains a major challenge in network science, which is  solved only in special cases. In a few cases exemplified by Turing patterns on graphs, collective states near the instability threshold are given by eigenvectors of the graph's Laplacian matrix \citep{nakao2010turing,hutt2022predictable}. Beyond such examples, the research is in an early phase where we still need systematic elements founding a broader theory \citep{prxPaper1, prxPaper2}. 

In several disciplines, a common approach is to study  relationships between structural and functional connectivity \citep{honey2009predicting,park2013structural,Aavik2014,messe2015closer,voutsa2021stylised}.
This approach has led to a rich phenomenology allowing important interpretations of data. However, due to its inherent link-by-link comparison of structure and dynamics, it is not well adapted for the study of \textit{collective} dynamical states.

We chose to consider Boolean attractors in threshold dynamics on signed directed graphs, and to investigate numerically under which topological conditions (involving e.g.  the cycle composition of the graph) eigenvectors can also predict collective states. 

Introduced in particular by Stuart Kauffman in 1969 \citep{kauffman1969homeostasis,kauffman1969metabolic} the Random Boolean Networks (RBN) model is a mathematical language to explore networks of interacting genes in a highly stylized form. In its original formulation with random Boolean update rules and random networks, where each node has the same number, $K$, of inputs, this model displayed statistical properties -- such as the transition from stable to chaotic behavior as a function of $K$ and the scaling of attractor numbers with network size \citep{mihaljev2006scaling} -- that fascinated statistical physicists and triggered substantial research activity in this field.

By including a threshold on the input in the dynamical rules ('threshold update rules', see below Section II A.), the predictive power of this stylized model for real-life biological systems became apparent \citep{li2004yeast,davidich2008boolean}, \citep[see also][]{bornholdt2005less}.

Similar discrete dynamics on graphs have been explored in the context of spin glasses \citep{bartolozzi2006spin}, cellular automata \citep{marr2005topology} and excitation spreading \citep{moretti2020link}. Technically, the threshold dynamics from \citep{li2004yeast} and \citep{szejka2008phase} belong to a category of totalistic cellular automata \citep[see][]{marr2009outer,marr2012cellular}. 

Due to its biological application and biological interpretability, Random Boolean Networks require signed directed graphs (in contrast to other types of discrete dynamics on graphs, like cellular automata). An edge $(v_1, v_2)\in E, v_i \in V$ of a graph $G(V,E)$ denotes the regulatory effect of the expression of a gene $v_1$ on the expression of another gene $v_2$. It is therefore condensed representation of an intricate set of biological processes (e.g.,  $v_1$ encoding a transcription factor, which has a binding site in the regulatory region of gene $v_2$).

The formal language of Boolean dynamics is now firmly established as one modeling approach in Systems Biology, among for example ODE models and metabolic flux models \citep{krumsiek2011hierarchical,choi2012attractor,saadatpour2013boolean,daniels2018criticality}. 
The Cell Collective database \citep{helikar2012cell} is devoted to this topic, as well as a segment of the BioModels database \citep{malik2020biomodels}. 

Threshold Boolean models have been criticised as models of biological systems \citep{zanudo2011boolean}, as they do not discriminate between functionally different nodes in the biological network, which are expected to follow distinct dynamical rules. 
While this criticism may also hold for gene regulatory networks \citep[see, e.g.][]{mozziconacci20203d}, the case of uniform (threshold-type) update rules is still a meaningful setting to further our understanding of these types of systems -- in particular the relationship between network architecture and the dynamics of gene activity --  as evidenced by the success of the approaches developed in \cite{li2004yeast,davidich2008boolean}. 
An early example of a detailed model with biologically motivated update rules designed individually for each node is presented in \citep{albert2003topology} \citep[see also][]{chaves2005robustness}.

Here we show (1) that signed directed random graphs differ strongly from each other in the correlation between eigenvectors of the graph and attractors under threshold Boolean dynamics, and (2) that the asymmetry of proportions of negative and positive cycles in the graph is a good predictor of this coupling of eigenvectors and dynamical attractors. In case of an over-representation of positive three-node cycles, a large percentage of attractors is structurally determined by the eigenvectors.

\section{\label{sec:methods}Methods}
\subsection{Boolean network model}
\begin{figure*}[t]
	\begin{center}
		\includegraphics[angle=-0,scale=0.45]{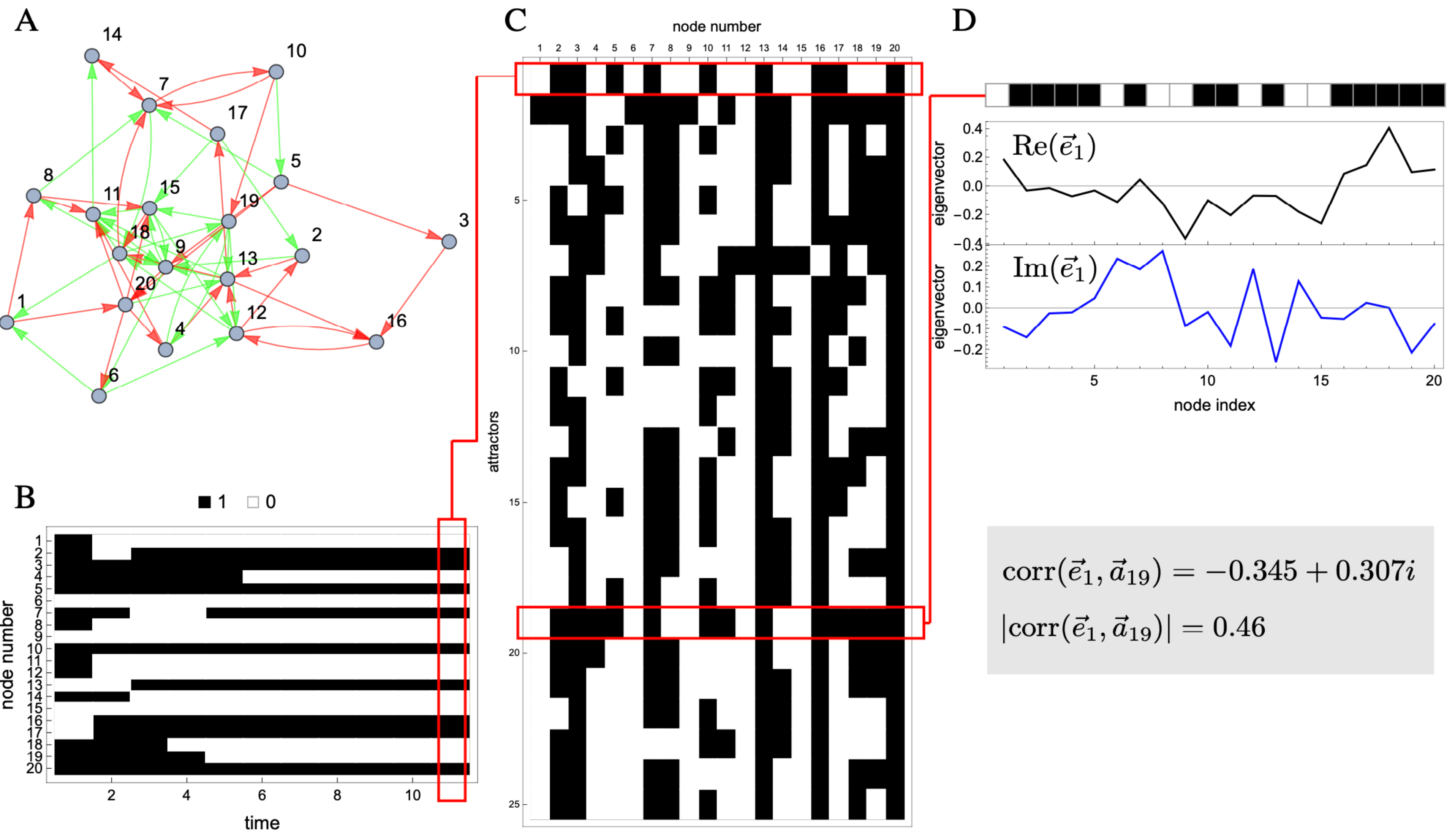}
		\caption{General outline of our investigation. (A) Example of a small stylized gene regulatory network with $n=20$ nodes and $m_+ = m_- = 30$ directed positive (activating) and negative (inhibitory) links shown in green and red, respectively. (B) Example of a simulated time course starting from random initial conditions on the network from (A) with a binary state space and the threshold-based update rule from \cite{li2004yeast}. The asymptotic state (fixed-point attractor) is reached after a short transient. (C) Set of 25 examples of fixed-point attractors for the network from (A). The attractor shown in (B) is given as the first entry in this set. (D) Top: Visual comparison of an attractor (attractor $\vec a_{19}$ from (C)) with one left eigenvector, $\vec e_1$, of the adjacency matrix of the network shown in (A) with the real and imaginary parts, $\mbox{Re}(\vec e_1)$, and $\mbox{Im}(\vec e_1)$, shown in separate panels under the attractor pattern. Bottom: Numerical result for the Pearson correlation coefficient between the Boolean attractor $\vec a_{19}$ and the (complex) eigenvector $\vec e_1$, $\mbox{corr}(\vec e_{1},\vec a_{19})$, together with the modulus of this quantity, $|\mbox{corr}(\vec e_{1},\vec a_{19})|$.  } \label{predFig1}
	\end{center}
\end{figure*}

The general design of our investigation is summarized in Figure \ref{predFig1}. 

Throughout the investigation the \textit{adjacency matrix}  $A$ of a directed network $G$ with nodes $\{1, 2, \dots n\}$ is defined as follows: $A_{ij}=0$ if there is no link from $i$ to $j$ in $G$, $A_{ij}=1$ if the link from $i$ to $j$ is activating, and $A_{ij}=-1$ if it is inhibitory. We consider only networks without self-loops, so $A_{ii}=0, i\in\{1, 2, \dots, n\}$.

The dynamical model used in this study is Boolean threshold dynamics. Every node $i\in \{1, 2, \dots n\}$ in the network $G$ has two possible states, $S_i = 1$ and $S_i = 0$. At every time step $t$, future node states $S_i(t + 1)$ are determined from present states $S_i(t)$ via the following update rule \cite{li2004yeast}: 
\begin{equation*}
	\begin{split}
		S_i(t + 1) = 
		\begin{cases}
			1, & \sum_{j=1}^{n}A_{ji}S_{j}(t) > 0 \\
			0, & \sum_{j=1}^{n}A_{ji}S_{j}(t) < 0 \\
			S_i(t), & \sum_{j=1}^{n}A_{ji}S_{j}(t) = 0.
		\end{cases}
	\end{split}
\end{equation*} 

Since the network is directed, its adjacency matrix is not necessarily symmetric, hence its right and left eigenvectors are different (and possibly complex). Given the expression $\sum_{j=1}^nA_{ji}S_j(t)$ in the above update rules, we are only considering left eigenvectors of $A$ in our analysis.

Note that in contrast to \cite{li2004yeast} we do not add inhibitory self-links to nodes, which only have positive input. In \cite{li2004yeast}  this was done with the goal of a quantitative comparison of the model's dynamics with (discretized) gene expression time series. For our statistical investigation, it is more appropriate to fully control the number of edges in the random graphs investigated here. 

The main quantity of our investigation is the \textit{predictability} of a given attractor $\vec{a}_{j}$. It is defined as the maximal (Pearson) correlation of this attractor with a left eigenvector $\vec{e}_{k}$:
\begin{equation}
	\pi (\vec a_{j})=\max_k|\mbox{corr}(\vec e_{k},\vec a_{j})|,
\end{equation}
where the maximum is taken over all (left) eigenvectors of the adjacency matrix of the network. 

In Figure \ref{predFig1}D eigenvector $\vec e_1$ has been selected for comparison, because for this eigenvector the correlation to the given attractor, $\vec a_{19}$, is maximal.

\subsection{Attractor predictability}
For the network shown in Figure \ref{predFig1}A we have obtained 181 distinct attractors by sampling 20.000 random initial conditions. The corresponding attractor predictabilities $\pi (\vec a _j)$ are shown as a histogram. \textit{Network-level attractor predictability} $\pi (G)$ of a network $G$ is now defined as the fraction of attractor predictabilities more than two standard deviations away from randomness, which is defined using the predictability distribution of randomized attractors. For the sake of convenience, we abbreviate the name of this quantity as \textit{network-level predictability} throughout the paper. For this particular example (the network shown in Figure \ref{predFig1}A) the network-level predictability is 0.57. 

\begin{figure*}[!htbp]
	\begin{center}
		\subfloat[][]{
			\includegraphics[angle=-0,scale=0.5]{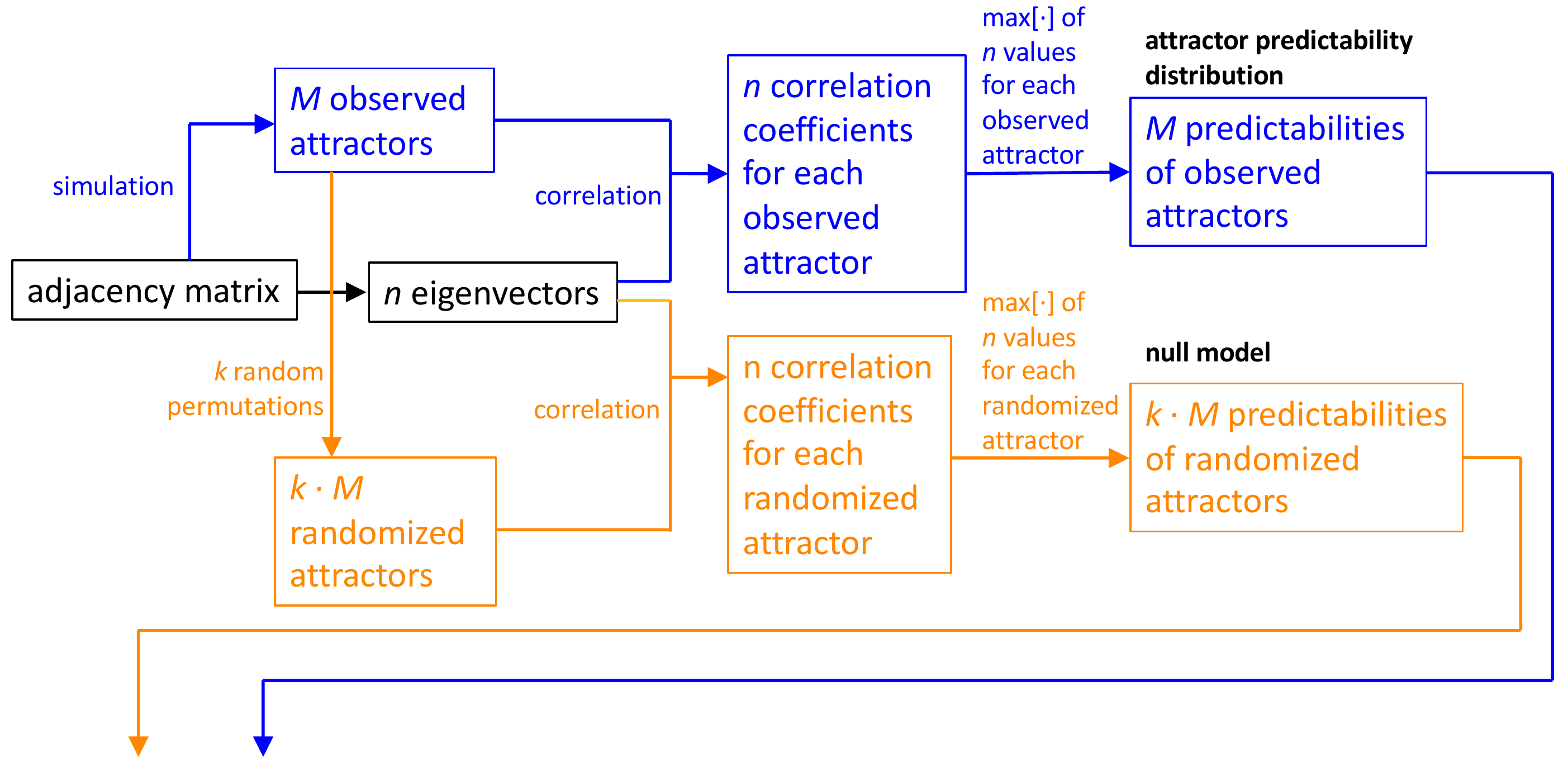}
		}
		\\
		\subfloat[][]{
			\includegraphics[angle=-0,scale=0.3]{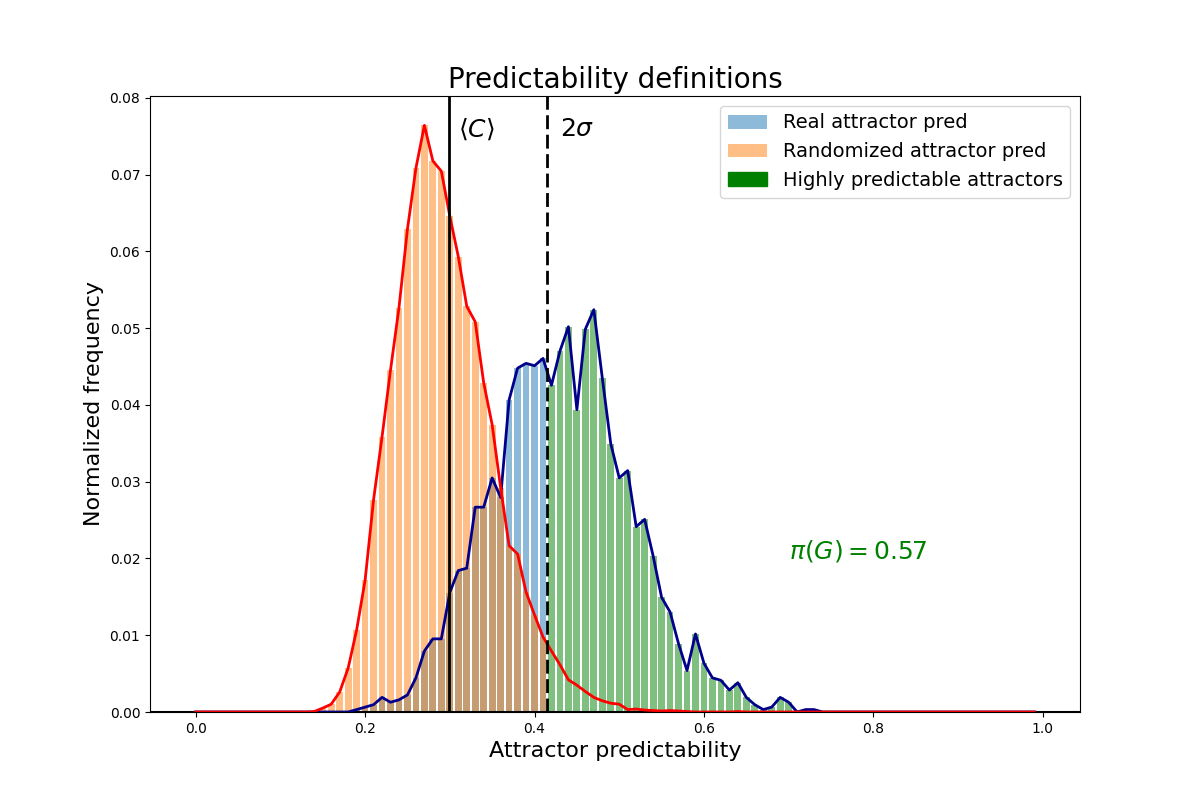}
		}
		\subfloat[][]{
		\includegraphics[angle=-0,scale=0.3]{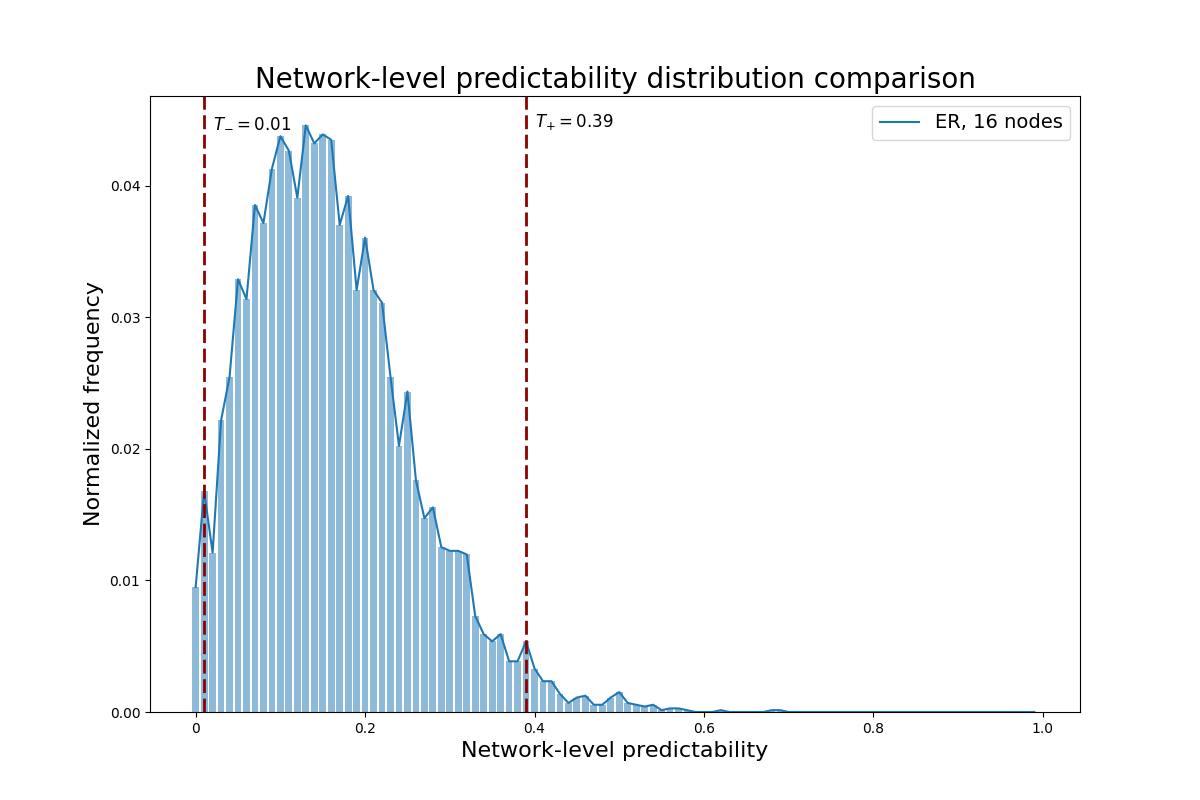}
		}
		\caption{(A) A schematic illustration of the computation of observed and shuffled attractor predictability distributions for a single network. Distinct attractors appear from many dynamical runs starting from different initial conditions. In practice, the number of dynamical runs is usually much larger than the number of observed attractors $M$, since distinct initial conditions often yield the same attractor. Random permutations of observed attractors which produce randomized attractors are chosen independently for each observed attractor. (B) Quantification of the network-level predictability $\pi (G)$ of an Erd{\"o}s-R{\'e}nyi (ER) network $G$ with $n=40$ nodes and $m_+=m_-=80$ directed positive and negative links (see Table ST1). Histogram of attractor predictabilities shown in blue, together with the corresponding histogram for randomized attractors (see \textit{Methods}) shown in orange. Cases contributing positively to $\pi (G)$ (i.e., those with $\pi (\vec a_j)> \langle C \rangle + 2\sigma $, with the average predictability of random attractors, $\langle C \rangle$, and its standard deviation $\sigma $) are highlighted in green. (C) Distribution of network-level predictability, together with the thresholds $T_-=0.01$ and $T_+=0.39$ for networks with high and low network-level predictability, respectively. $T_+$ was chosen as the largest value (up to two decimal paces) that labels at least 2\% of all synthetic ER networks described in Section III B. as those with high network-level predictability, and $T_-$ was chosen similarly with the same 2\% threshold in mind. Only networks with at least 50 distinct attractors enter this histogram.} \label{predFig2}
	\end{center}
\end{figure*}

A summary of the pipeline for finding predictability distributions of observed and shuffled attractors is shown in Figure \ref{predFig2}A. Figure \ref{predFig2}B illustrates the definition of network-level predictability.  

\subsection{Network generation model for varying asymmetry in positive and negative directed cycles}
In a stylized regulatory network a directed cycle is said to be \textit{positive} if it contains an even number of inhibitory links (possibly zero), and is said to be \textit{negative} otherwise. Adopting the standard definition of an asymmetry (see, e.g., \cite{kosmidis2020chromosomal}), the \textit{asymmetry of a network $G$ for cycles of length $k$}, $A_{k}(G)$, (brief: $k$-cycle asymmetry or, for the case $k=3$, dominantly considered in the following discussion, simply cycle asymmetry) is then defined as
\begin{equation*}
	\begin{split}
		A_{k}(G) = \dfrac{N_{k,+}(G) - N_{k,-}(G)}{N_{k,+}(G) + N_{k,-}(G)},
	\end{split}
\end{equation*} 
where $ N_{k,+}(G) $ and $N_{k,-}(G)$ denote the numbers of positive and negative cycles of length $k$ in $G$, respectively.

To create a directed network of $n$ nodes with $m_+$ positive and $m_-$ negative links, which has a value of 3-cycle asymmetry close to $a_{target}$, we first generate a random directed Erd\H{o}s-R{\'e}nyi network with exactly $m_+ + m_-$ unsigned links. Then we iterate through all of its 3-cycles in random order, allocating signs of links in those cycles in a way which brings 3-cycle asymmetry closer to $a_{target}$ whenever possible. The algorithm also strives to maintain a balance between the number of different $\pm1$ 3-cycle compositions in the resulting network. As the final step, signs of links that are not part of any 3-cycle are chosen randomly, subject to $m_+$ and $m_-$ constraints. 

The sets of high- and low-asymmetry networks used in this study contain 10.000 samples each, with $n=16$, $m_+=m_-=32$ and $a^{high}_{target}=0.8, a^{low}_{target}=-0.8$, respectively. They have average 3-cycle asymmetries of $0.79308$ and $-0.80284$, with the standard deviation of less than $0.05$.

\subsection{Generation of signed directed regular random networks}
For a specific network node, we denote by $k_{out}^{+}$, $k_{out}^{-}$, $k_{in}^{+} $ and $k_{in}^{-}$ the number of links of corresponding sign and direction. A signed directed regular network of degree $k$ is a directed signed network where each node has $k_{out}^{+}=k_{out}^{-}=k_{in}^{+}=k_{in}^{-}=k.$

In order to generate a signed directed random regular network of degree $k$ on $n$ nodes, we first create an undirected unsigned connected random regular network of degree $4k$. Next, we iterate through the nodes in random order and distribute link directions randomly, compliant with degree constraints $k_{out}=k_{out}^{+} + k_{out}^{-}=2k$ and $k_{in}=k_{in}^{+} + k_{in}^{-}=2k$. Afterwards, subsets of $k$ positive outgoing links and $k$ positive incoming links are randomly chosen for each node, which fully determines signs of the remaining links. During those two steps, only random assignments of directions and signs, which do not lead to immediate conflicts (such as making a link whose sign is fully determined by the process of elimination with respect to one node violate degree constraints of another node) in the network, are performed. Finally, in order to avoid biasing the output towards specific topological configurations, direction- and sign-preserving  pairwise switch randomization is run on the network for a large number of steps.

In this study, we use a set of $10.000$ connected signed directed regular random networks of degree 2 with $16$ nodes. The numbers are chosen for consistency with the ER dataset (see Section III B.), since resulting networks also have $64$ links in total. 

We define sets of regular networks with high and low network-level predictability by using the 2\% quantile-motivated thresholds $T_+: \pi(G) \geq 0.66, T_-: \pi(G) \leq 0.13$,

\subsection{Numerical experiments}
\subsubsection{Standard settings}

Our main example will be stylized gene regulatory networks - random Erd{\"o}s-R{\'e}nyi networks with $n=16$ nodes, $m_+=32$ directed positive and $m_-=32$ directed negative links. We created a database by running simulations on 50.000 such networks of this size and for each network computing all distinct attractors from the complete set of 65.536 possible initial conditions.

\subsubsection{Filtering criteria for biological networks}

We analyze network-level attractor predictability for the 78 networks of models from the Cell Collective database \cite{helikar2012cell}, as well as the 9-node network from \cite{zhang2014stem} and 6 biological networks from \cite{tripathi2020biological}. In order to have meaningful structural information in these networks, we apply filtering criteria on connectivity (weak connectivity in [0.15,1]), number of observed fixed-point attractors (at least 10), number of cycles (at least 1 cycle of length 3 or 4) and the ratio of positive and negative links (between 0.3 and 0.65). This leads to 15 networks included in our analysis.

\subsubsection{GeneNetWeaver experiments}
Simulations with GeneNetWeaver \citep{schaffter2011genenetweaver} were performed in the following way: We considered the averaging GeneNetWeaver model (see Supplements) with measurement noise, as well as a stochastic version with both intrinsic and measurement noise using GeneNetWeaver default parameters. We used 400 stylized gene regulatory networks from the database described in standard settings, and calculated GeneNetWeaver attractors for each of them from 250 random initial conditions.

\subsubsection{Analysis of small networks}

We define in-degree asymmetry of a node $k$ in a network $G$ as \begin{equation*}
	\begin{split}
		a_{G}(k) = \dfrac{n_{+}^k - n_{-}^k}{n_{+}^k + n_{-}^k},
	\end{split}
\end{equation*} 
where $ n_{+}^k $ and $n_{-}^k$ denote the numbers of positive and negative links incoming to node $k$, respectively.

In Figure \ref{additional_figure_1}B a threshold of $\pm0.6$ is used to define 3-cycle asymmetry extremes. The other quantity studied in that experiment is the number of network nodes with extreme in-degree asymmetry, with node $k$ considered extreme if $|a_{G}(k)| > 0.6$. Due to network architecture, the number of such nodes has 3 possible values: 3, 4 or 5.  All 128 link sign assignments are analyzed, with no constraints on the fraction of positive links. 

The algorithm employed in Figure \ref{additional_figure_1}D performs a single link sign swap on every iteration. To provide a smoother trajectory of network-level predictability, changes 3-cycle asymmetry as gradually as possible in the desired direction, while also trying to affect the smallest number of cycles. Among candidates that satisfy those two costraints, selection of link sign swap is random, to eliminate possible bias. 
\section{\label{sec:results}Results}

\subsection{Attractor predictability as a network-level property}

We investigate network-level predictability in three ways: (1) We create a database of ER networks with high and low network-level predictability values and discover that the two populations differ systematically in cycle asymmetry. To ensure that this effect is not a consequence of node degree, we generate a large set of signed directed \textit{regular} random networks and use it to validate our findings (Section III B.). 
(2) Using an algorithm which allows us to systematically vary the cycle asymmetry of a network, we validate the statistical significance of cycle asymmetry's ability to discriminate high and low network-level predictability (Section III C.). (3) We study small graphs of only few cycles to gather some analytical insight about the way attractor-eigenvector relationship emerges and how this relationship changes as a function of cycle sign (Appendix B).

\subsection{Topological properties of networks with high and low attractor predictability}

With the basic outline of our approach given in Figure \ref{predFig1} and the key quantity we analyze -- network-level predictability of a network -- illustrated in Figure \ref{predFig2}B, we can now turn to the quantitative investigation of a larger ensemble of graphs. The settings for those numerical experiments are described as "standard settings" in Methods.  Figure \ref{predFig2}C shows the distribution of network-level predictability, together with the thresholds we used to define the subsets of networks with high and low network-level predictability, respectively. These two sets now allow us to investigate topological differences between networks with extreme values of network-level predictability. 

\begin{figure}[t]
	\begin{center}
		\subfloat[][]{\includegraphics[angle=-0,scale=0.3]{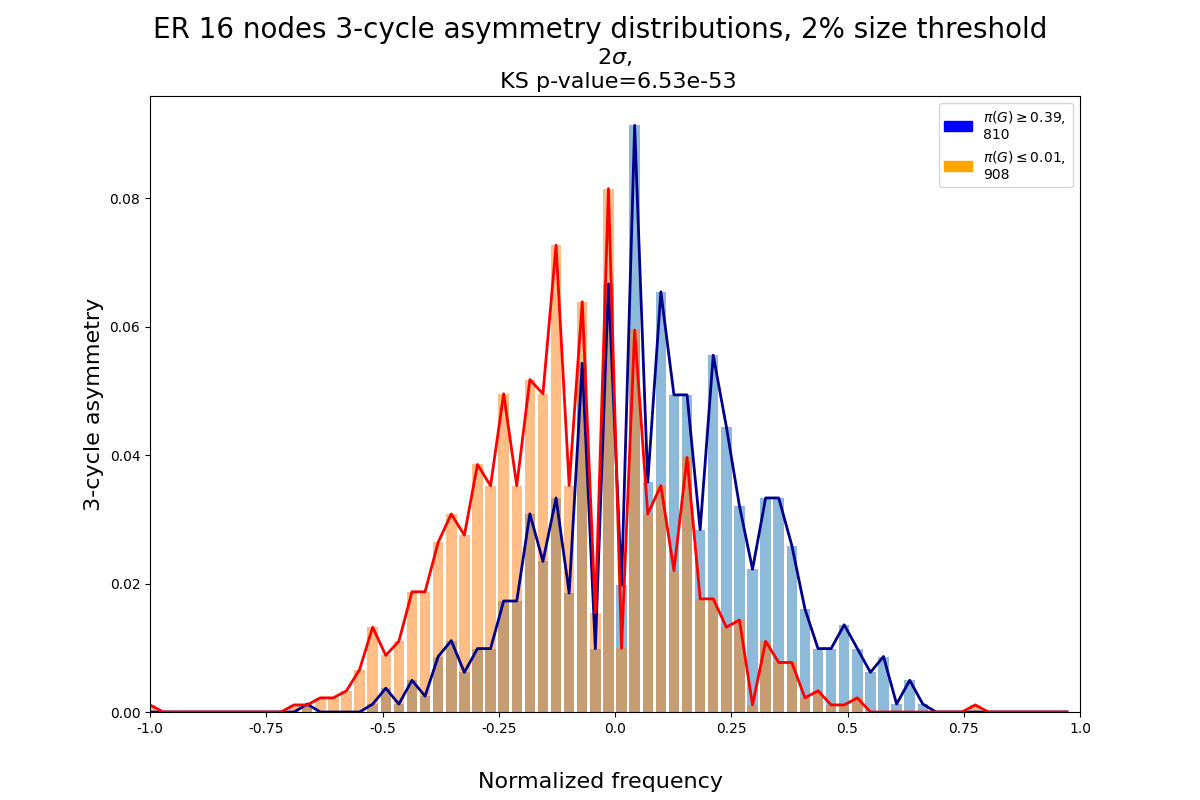}}\\
		\subfloat[][]{\includegraphics[angle=-0,scale=0.3]{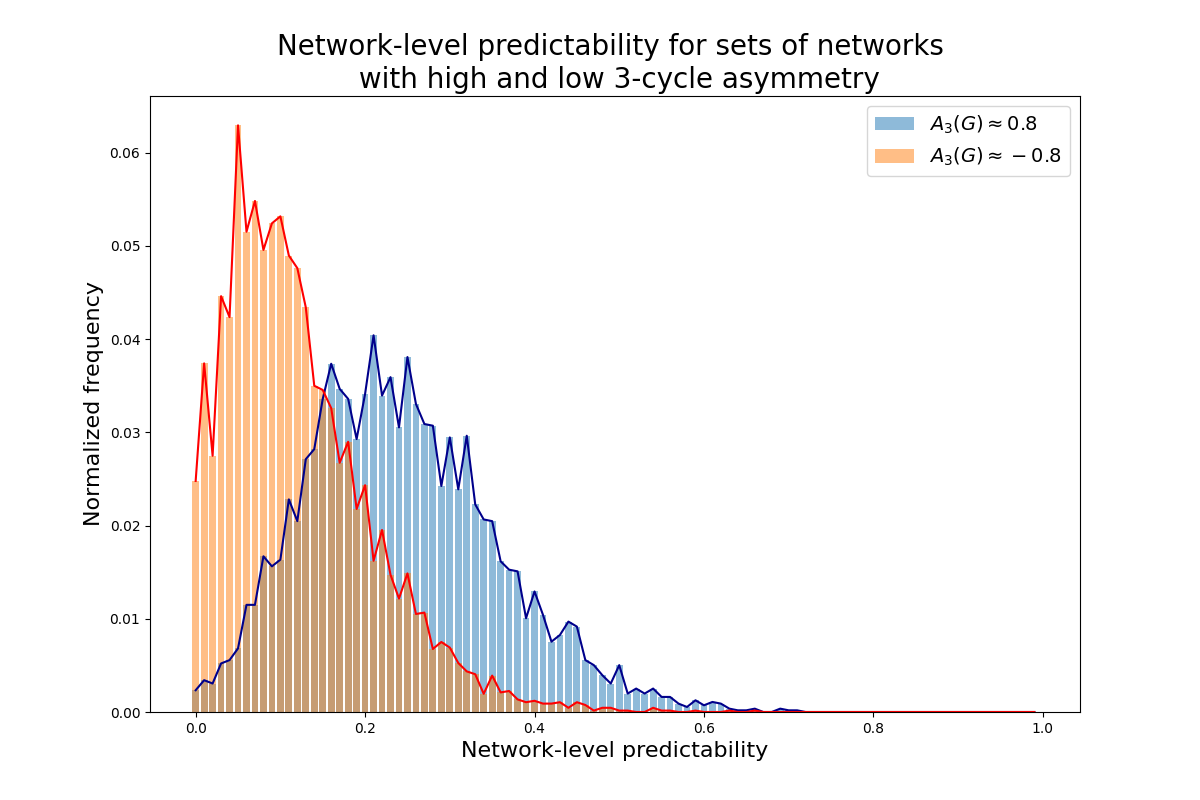}}
		\caption{(A) Histogram of 3-cycle asymmetry distributions for sets of networks with high (blue) and low (red) network-level predictability, defined using 2 standard deviation threshold (see \ref{predFig2}B). KS p-value corresponds to the p-value obtained from two-sample Kolmogorov–Smirnov test (B) Histograms of network-level predictability for networks with many negative cycles (negative 3-cycle asymmetry; orange) and many cycles (positive 3-cycle asymmetry; blue).} \label{predFig4}
	\end{center}
\end{figure}

While most of the usual topological quantities (degree distributions, centrality measures, assortativity, etc.)  were surprisingly uninformative in discriminating these two sets of networks (see Figures S11-S13), striking topological differences become visible on the level of small directed cycles, if the signs of links which form the cycle are taken into account. Figure \ref{predFig4}A shows the asymmetry of positive and negative 3-cycles (see \textit{Methods}) for the two sets of graphs. Those sets display a  difference in observed cycle asymmetry that is significant with a $p$-value of $6.53\cdot 10^{-53}$ according to a two-sample Kolmogorov-Smirnov test (see \ref{predFig4}A). Furthermore, while using a 2-standard deviation network-level predictability threshold (see \ref{predFig2}B), we see that networks with high network-level predictability assume relatively high ($A_3(G) \geq 0.4$) positive values of 3-cycle asymmetry 662\% more often than those with low network-level predictability. Conversely, networks from the latter set exhibit a high ($A_3(G) \leq -0.4$) fraction of negative 3-cycles 462\% more often than those from the former one. 

To ensure that observed topological difference between sets of networks with high and low network-level predictability is not a consequence of the properties of degree distribution, it is necessary to eliminate it as an intervening factor. We approach the problem by constructing signed directed regular random networks of the same size and connectivity(see \textit{Methods}) and studying this new dataset with the same methods that we describe for ER networks in this section. 

 We discover that asymmetry of positive and negative 3-cycles discriminates between networks with high and low network-level predictability (see Methods) as well (see Figure S7).

\subsection{Network-level predictability for networks with varying cycle asymmetry}
The main topological difference between highly predictable and poorly predictable networks, so far, has been observed on a statistical level using a large set of random networks. 

Next, in order to assess whether indeed this topological difference can also be causally linked to network-level predictability, we create networks with controlled asymmetry of positive and negative cycles (see \textit{Methods}) and analyze the network-level predictability for sets of networks with high and low cycle asymmetry, respectively. We discover that networks where the majority of cycles are positive exhibit predictability values between 0.3 and 0.6 614\% more often than their counterparts with low cycle asymmetry. Conversely, networks with low cycle asymmetry have network-level predictability in [0, 0.15] 243\% more often than those with high cycle asymmetry (see Figure \ref{predFig4}B).

\subsection{Continuous dynamics and noise}
An important question is, whether network-level attractor predictability goes beyond the specific Boolean dynamical model and extends to, for example, continuous gene expression dynamics simulated via differential equations. In order to address this point, we use the GeneNetWeaver tool \citep{schaffter2011genenetweaver}, a model based on ordinary differential equations (in the case of no noise or just measurement noise) or stochastic differential equations (in the case of intrinsic noise). GeneNetWeaver has served as a generator of synthetic gene expression data for several DREAM gene network inference competitions \citep{marbach2010revealing,marbach2012wisdom}. 
\begin{figure}[t!]
	\begin{center}
		\includegraphics[angle=-0,scale=0.3]{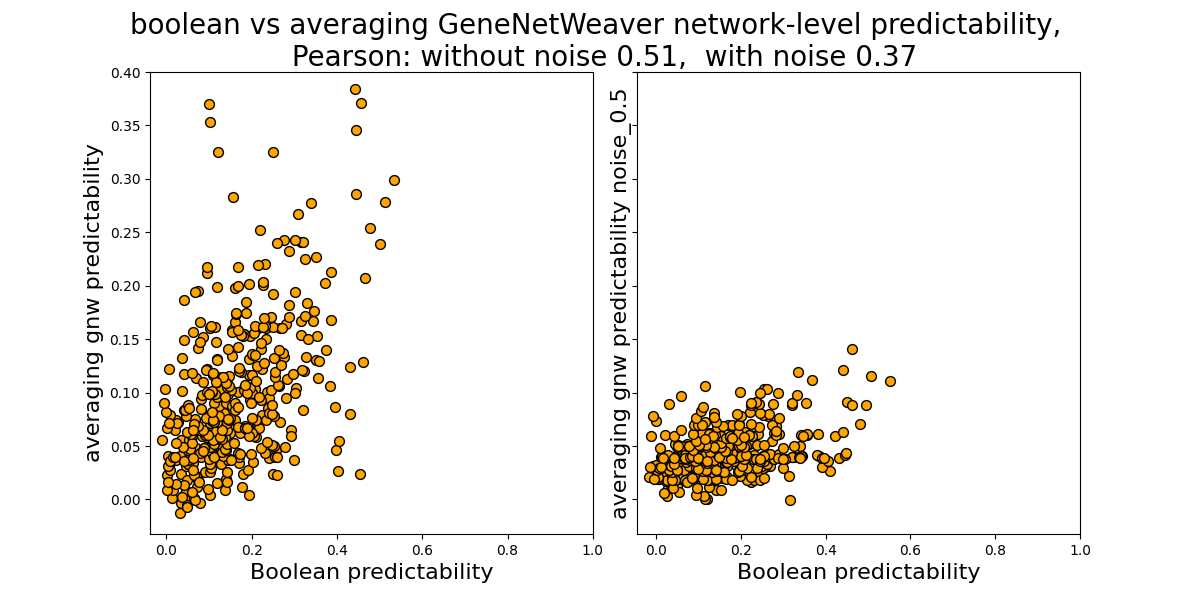}
		\caption{Comparison of network-level attractor predictabilities derived from the Boolean model with those derived from the GeneNetWeaver data simulator. (A) GeneNetWeaver simulation with measurement noise but no intrinsic noise. (B) GeneNetWeaver simulation with both measurement noise and intrinsic noise. Default values are used for measurement noise, while intrinsic noise corresponds to noiseCoefficientSDE=0.5 in GeneNetWeaver settings (10 times the default value)}\label{additional_figure_2}
	\end{center}
\end{figure}
Note that GeneNetWeaver has by default a multiplicative activation term \citep{schaffter2011genenetweaver}. For our simulations we rewrote this as an additive activation function to be comparable to the threshold dynamics used in our Boolean model. A summary of these activation functions and results for the original GeneNetWeaver model are given as Supplementary Information (see Figure S10 and relevant text). We simulate steady states for several initial conditions (see Methods) and we compute, as in the case of Boolean dynamics, the correlation of these asymptotic states with the eigenvectors of the graph and thus obtain network-level predictabilities. 

Figure \ref{additional_figure_2}A shows that in spite of the markedly different nature of the two models the predictabilities show a clear positive correlation. This positive correlation persists even in the presence of intrinsic noise (Fig. \ref{additional_figure_2}B). 

\subsection{Small networks}

\begin{figure}[t!]
	\begin{center}
		\includegraphics[angle=-0,scale=0.3]{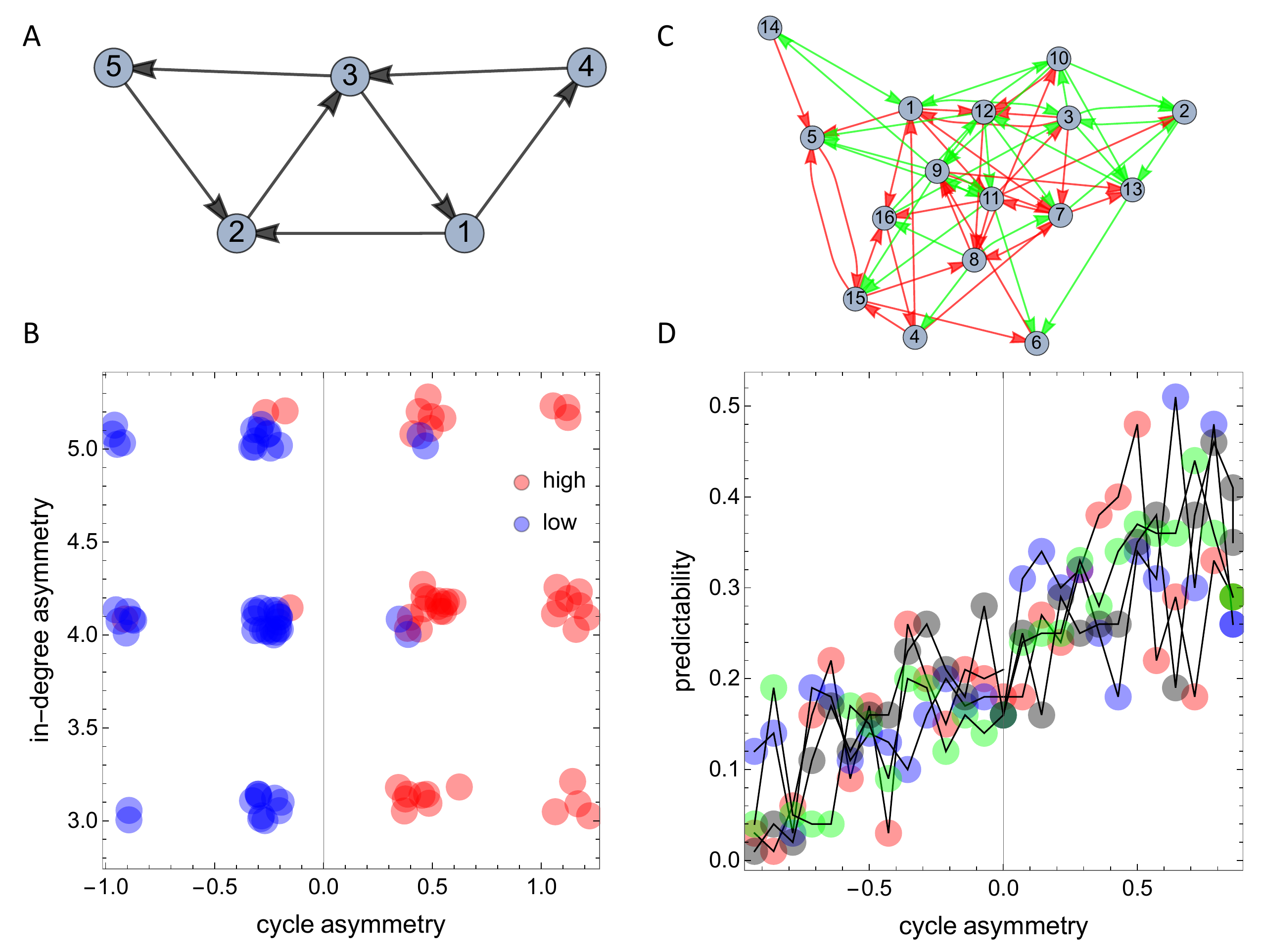}
		\caption{Case study on a small graph with three 3-cycles and design of highly predictable networks. (A) Graph with $N = 5$ nodes and $M = 7$ edges, arranged to form three 3-cycles. (B) Distribution of high (red) and low (blue) predictability in the plane of cycle asymmetry and extreme in-degree asymmetry node count. For defining such extreme nodes, in-degree asymmetry threshold of $0.6$ has been employed. Predictability thresholds $>0.2$ and $<0.1$ have been used to define high and low predictability, respectively. Note that a small random shift of points has been applied for visual clarity. (C) Initial graph used for the design of highly predictable networks ($N = 16$, $M = 64$, number of 3-cycles $N_C = 28$. (D) Changes of predictability under step-wise cycle sign switches. Starting from the graph displayed in (C) we gradually increase the number of positive cycles (to the right) or the number of negative cycles (to the left) and observe the corresponding change in network-level predictability. Four independent trajectories are shown in different colors. 
		} \label{additional_figure_1}
	\end{center}
\end{figure}

In order to develop some intuition about the observed statistical association between network-level predictability and network architecture, we study a minimal example consisting of five nodes and seven edges arranged to form three overlapping triangles, as depicted in Figure \ref{additional_figure_1}A. Starting from this template we create all possible signed graphs and study the two topological quantities, cycle asymmetry and the number of nodes with extreme in-degree asymmetry (see Methods for definitions) as candidates for explaining network-level attractor predictability. Structurally, this procedure leads to four values of 3-cycle asymmetries (all three cycles negative, $A=-1$, two cycles negative, one positive, $A= -1/3$, etc.) and three values of extreme in-degree asymmetry node count (see Methods for details). For all signed graphs we compute attractor predictability as previously described and assign categories of high and low predictability via thresholds.   Figure \ref{additional_figure_1}B shows the distribution of these two categories in the plane spanned by the two topological quantities. Cases of high and low predictability are almost perfectly separated along the cycle asymmetry axis. The few remaining exceptions are situated at high in-degree asymmetry. We verified that this general picture does not change qualitatively when varying these choices of thresholds used for predictability categories within reasonable ranges. 
Supplementary Information shows a distribution of network-level predictabilities (see Figure S16).

Summarizing we see that even in these small stylized graphs, high positive cycle asymmetry leads to high predictability, with a slight secondary influence from in-degree asymmetry.

As a next step, we can attempt to \textit{design} highly predictable networks by adjusting the cycle content. In order to do this, we create a random directed ER graph, randomly distribute an equal number of positive and negative signs on the edges (with the subsidiary constraint that the initial 3-cycle asymmetry is zero) and then switch signs (conserving the balance of positive and negative edges) such that 3-cycles are iteratively turned into positive and into negative cycles, respectively. The initial graph is given in Figure \ref{additional_figure_1}C. Examples of the corresponding trajectories of predictability as a function of cycle asymmetry are shown in Figure \ref{additional_figure_1}D.

\subsection{Application to biological examples}

\begin{figure}[!htbp]
	\begin{center}
		\includegraphics[angle=-0,scale=0.3]{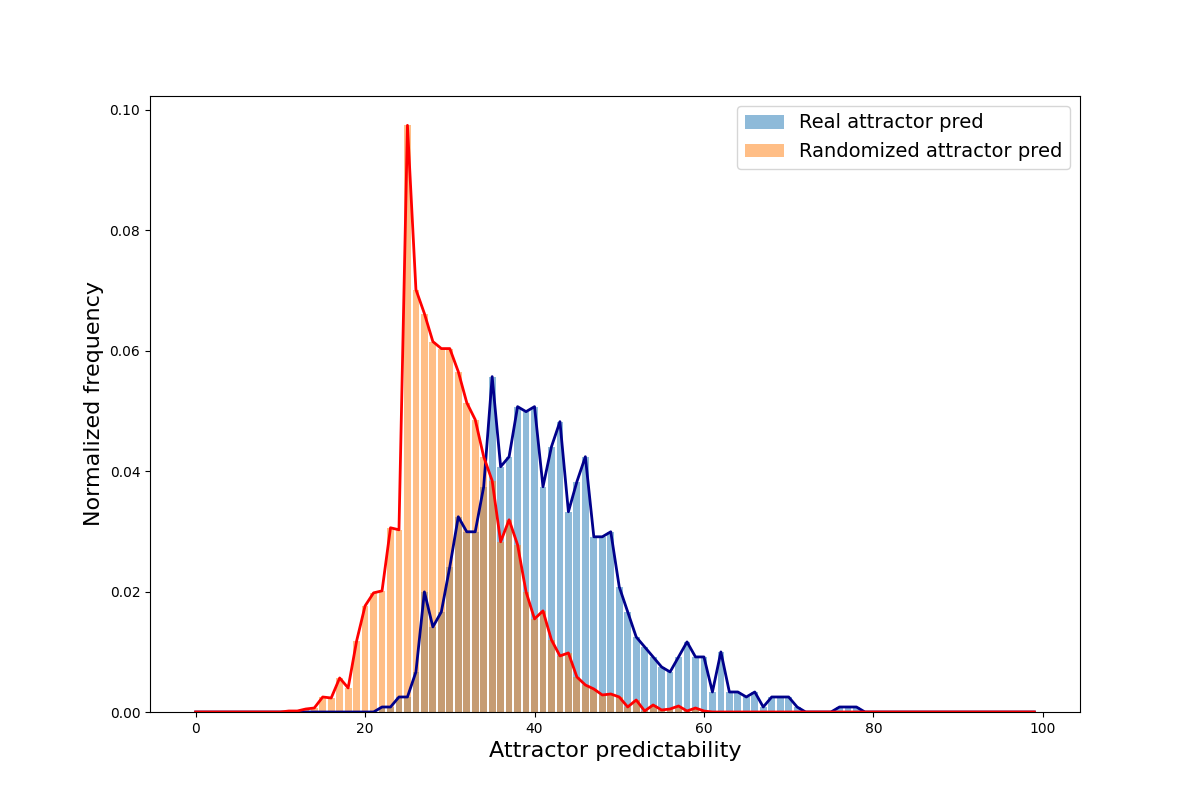}
		\caption{Normalized predictability histograms of \textit{Tumour cell invasion and migration} model network attractors and their shuffled versions. For each attractor (except the constant zero vector) five random permutations are considered. The network has 3-cycle asymmetry of $A_3(G)=0.53$, and network-level predictability of $\pi(G)=0.36$ when the 2-standard deviation threshold is used (see \ref{predFig2}B).} \label{TumourPredictability}
	\end{center}
\end{figure}

To assess how relevant the notion of predictability is for real-life biological networks we apply our methods to networks of models from Cell Collective database \cite{helikar2012cell}, as well as networks studied in \cite{zhang2014stem} and \cite{tripathi2020biological} (see Methods). Figure \ref{TumourPredictability} shows an example of attractor predictability distribution for Tumor Cell Invasion and Migration network. 

In line with studying attractor predictabiltiy as a network level property, we analyze its connection to 3-cycle asymmetry for a subset of networks obtained after applying a set of filtering criteria to the collection described above (see Methods). Table \ref{table1} shows that this topological property still discriminates between networks with high and poor network-level predictability.
\newline

\begin{center}
	\begin{table}[!hbtp]
		\caption{ Predictability $\pi(G)$, asymmetries $A_3(G)$ and $A_4(G)$ and agreement with theory for biological networks. Thresholds for high and low cycle asymmetry are $\pm0.25$. Low and high network-level predictability thresholds are $\pi(G) < 0.1$ and $\pi(G) > 0.3$.  Agreement is assigned as follows: if both $A_3(G)$ and $A_4(G)$ are below $-0.25$ for low or above $0.25$ for high network-level predictability, the results is $++$. If the previous condition holds for either $A_3(G)$ or $A_4(G)$, and the other asymmetry is between the thresholds, the result is a $+$. If network-level predictability and one of asymmetry values are extreme in the opposite sense, and the other asymmetry value is between the thresholds, the result is a $-$. Results of $-+$ and $--$ are defined similarly, and are not present in the table. \\}\label{table1}
		\resizebox{0.47\textwidth}{!}{
		\begin{tabular}{|l|c|c|c|c|}
			\hline
			\textbf{Model name}  & $\pi(G)$ & $A_3(G)$ & $A_4(G)$ & \textbf{Agreement}\\
			\hline
Mammalian Cell Cycle 2006 &0.01&0.20&-0.25&
\\ \hline Budding Yeast Cell Cycle 2009 &0.07&-0.33&-0.76&++
\\ \hline CD4+ T Cell Differentiation and Plasticity &0.07&0.33&0.04&-
\\ \hline Oxidative Stress Pathway &0.08&-1.00&-1.00&++
\\ \arrayrulecolor{orange}\hline \arrayrulecolor{black} Budding Yeast Cell Cycle &0.11&1.00&-0.27&
\\ \hline Arabidopsis thaliana Cell Cycle &0.15&-0.52&0.00&
\\ \hline emt26network.csv&0.16&0.80&0.93&
\\ \hline Lac Operon &0.21&nan&1.00&
\\ \arrayrulecolor{blue}\hline \arrayrulecolor{black} T-LGL Survival Network 2011 Reduced Network &0.31&-0.33&0.00&-
\\ \hline B cell differentiation &0.32&1.00&1.00&++
\\ \hline gonadalsexdet.csv&0.34&0.64&0.62&++
\\ \hline Tumour Cell Invasion and Migration &0.36&0.53&0.27&++
\\ \hline sclcnetwork.csv&0.78&0.50&0.37&++
\\ \hline gastricnetwork.csv&0.79&0.42&0.30&++
\\ \hline Aurora Kinase A in Neuroblastoma &0.95&0.20&0.43&+
\\ \hline 
		\end{tabular}}	
	\end{table}
\end{center}
Even though the overall agreement between cycle content and network-level predictability is not as clear as in our investigation of random graphs, the association is still quite visible in Table \ref{table1}. Note that in spite of the selection criteria we impose (see Methods), the real networks tend to be rather sparse. As a consequence, the numerical values of cycle asymmetry may not always be reliable, due to the small number of cycles in the graph (cycle statistics are provided in Table ST2) and, in contrast to random graphs, where the cycle asymmetry derived from 3-cycles was by far the most dominant topological feature, here both 3-cycles and 4-cycles need to be taken into account. 

\section{\label{sec:discussion}Discussion}

As with any systematic investigation of this type, our analysis contains several tunable parameters and one needs to understand, how the observations made here depend on these parameters. 

The main parameters are: 

(1) The connectivity of the network. There the factors limiting the range of this parameter are connectedness of the network (distinguish between strongly and weakly connected graphs) at the lower end and the rapid decline of the number of distinct attractors on the upper end (see Figure S2). Nevertheless, we found that our results remain valid for a range of weak connectivity values in [0.15, 0.6] (see Figure S3) 

(2) The asymmetry of positive and negative links. The main part of our investigation has been performed with equal numbers of positive and negative links. In order to gain some insight in how the results depend on the positive-negative link asymmetry, we first look at an estimate of the average number of attractors as a function of this asymmetry (see Figure S1) As expected, the number of attractors decreases when the fraction of positive links is very high. This does not happen for very low fraction of positive links, but those values are excluded from consideration to eliminate possible bias they introduce into the dynamical behavior of the system.

This delineates a range of asymmetries for which predictability can be meaningfully investigated. 

Next we can look at the average predictability as a function of asymmetry between positive and negative links. This can be done in two ways: (i) on the level of attractors, studying the distribution of the attractor-eigenvector correlations (in comparision with a null model of shuffled attractors) as a function of the positive-negative link asymmetry; (ii) on the level of networks, where we can look at the network-level predictability (i.e., the fraction of attractors with value of predictability more than two standard deviations way from the average  of random predictions). Based on our analysis, results of this study do not change qualitatively when asymmetry of positive and negative links lies within this range (see Figure S4).

(3) The last parameter to be addressed is the most challenging one: network size. A full enumeration of all initial conditions quickly becomes unfeasible when we depart from network sizes explored here. As this parameter increases, random sampling covers an ever smaller percentage of initial conditions, thus biasing the investigation towards attractors with large basins. 

In order to see, whether this bias towards attractors with a large basin has already a substantial effect in the case of the network sizes under investigation here, we also conduct an investigation of larger networks with attractor sets drawn from randomly sampled initial conditions. Supplementary experiments indicate that our results remain valid for networks of size 40 with 50.000 random initial conditions each (see Figure S6), and although sampling of large basin attractors tends to somewhat amplify extreme cases of network-level predictability, results of this study are not affected qualitatively (See Figure S9). In addition, we verified that even if complete enumeration of initial conditions is impossible, network-level predictability is robust with respect to different samplings of initial conditions performed on the same network (see Figure S8). We furthermore checked, that our key results do not depend on the exact choice of the minor technical parameters of our investigation, namely the number of sampled initial conditions (see Figure S5), the attractor predictability threshold (i.e., two standard deviations away from randomness, or quantile-based definitions; see Figures S14, S15), and the thresholds used to identify highly predictable  and poorly predictable networks (see \ref{predFig4}A). 

To summarize the above discussion, we expect results of our study to remain valid at least for the following parameter ranges: weak connectivity in [0.15, 0.6] (for small networks lower bound is greater since it is difficult to observe a meaningful amount of cycles at low connectivity there), proportion of positive links in [0.3, 0.65], network size between 4 and 40 (but likely larger networks as well, since random sampling strategy does not change), predictability threshold between 1 and 3 standard deviations away from randomness, and a threshold between 0.01 and 0.1 of total dataset size for composing sets of highly and poorly predictable networks. A meaningful lower bound on the number of observed distinct attractors can be chosen by analyzing the distribution of those numbers for fixed connectivity, positive/negative link ratio and network size. For $n=16$ with $m_+=m_-=32$, which has been dominantly considered in this study, we require at least 50 distinct attractors. 

In \cite{tripathi2020biological} the parallel of the Boolean model to an asymmetric spin glass model is used to define frustration (in the sense of \cite{anderson1978concept}). It is argued that the level of frustration may indeed be a non-random feature of biological regulatory networks. Very much in the light of \cite{tripathi2020biological}, our research is motivated by two questions, which to date are not fully answered: (1) Which constraints does the gene regulatory network impose on attractors? (2) Which non-random features do gene regulatory networks have and how are they of relevance for their biological function?

\section{\label{sec:conclusion}Conclusion}
We provide a way of quantifying predictability of collective dynamical states by eigenvectors. And we offer heuristics, when networks have a high predictability (a strong relationship between eigenvectors and attractors) and when they have a low predictability. We also show that the link between eigenvectors and attractors are not universal or omnipresent, but the details of the underlying network matters, in a systematic and predictable way. 

In particular, our findings install some confidence that genetic programs and patterns of gene activity are tighly constrained by the architecture of the underlying regulatory network. Our findings also highlight the importance of accumulating enough knowledge about regulatory machineries. 

It is an obvious next step to extend these findings to real gene expression patterns. Such an application is limited at the moment by two major factors: (1) the incompleteness of our knowledge of gene regulatory networks, even for the simplest organisms. Today, the most
comprehensive transcriptional regulatory network is the one stored in RegulonDB \cite{huett_lesne2021}. However, our knowledge about most interactions in this network is partial. In particular, the regulation of many promoters is presently unknown \cite{regulon2016}. Transition to multicellular organisms further complicates research, because the data becomes less and less complete \cite{huett_lesne2021}; (2) the fact that real gene expression patterns do not exclusively arise from the action of the transcriptional regulatory network, but are subject to a multitude of other biological mechanisms; in the case of bacterial gene regulation, for example, the influence of chromosomal spatial organization has been explored in much detail over the last years \cite{eda2021}. 

In our investigation we focus on the predictability of attractors from eigenvectors of the underlying regulatory network. Noise, a topic of high relevance in studying gene expression patterns \citep{thattai2001intrinsic,chalancon2012interplay,zhang2014stem}, therefore is less important in our investigation, as it can be thought of as an influencing factor of the transient \textit{towards} an attractor and consequently as a selective mechanism affecting only the statistics of the attractors. However, the robustness of an attractor with respect to (e.g., update timing) noise \citep{klemm2005topology,bornholdt2008boolean,braunewell2009reliability} is an important characteristic that deserves further investigation. In future work we intend to explore, whether highly predictable attractors are more likely robust than poorly predictable attractors, which would suggest that the robustness of a dynamical state would be structurally determined. Some evidence of such structural effects of robustness is provided in \cite{klemm2005topology}. 

More generally, our work suggests to resort to eigenvectors to predict collective dynamical states, substantially beyond the well-known example of Turing patterns. So far, our results are a set of statistical observations. However, this statistical approach appears very powerful to unravel  unexplored relationships between network architecture and collective states. The challenge is now open to elaborate definite theoretical statements in  specific situations from these broad statistical relationships.

\begin{acknowledgments}
M. T. Hütt  acknowledges support by Volkswagen Stiftung (grant number 9A174). M. T. H\"utt thanks LPTMC (Paris) for hospitality and the Physics Institute of CNRS (French National Center for Scientific Research) for funding his stays, during which part of this work has been performed. 
\end{acknowledgments}

\appendix

\section{Illustration of the numerical procedure for a small graph}
\begin{figure}[!htbp]
	\begin{center}
		\includegraphics[angle=-0,scale=0.3]{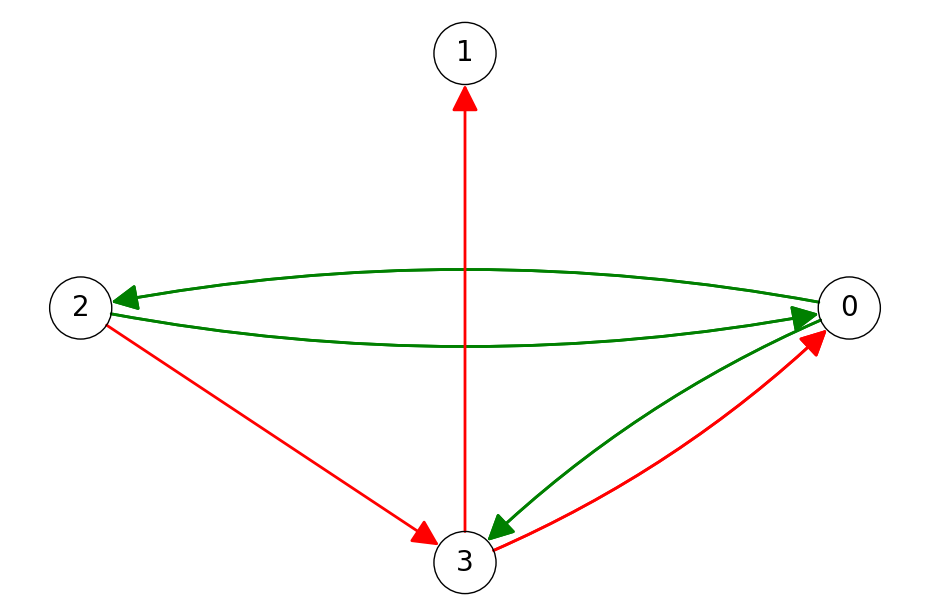}
		\caption{The network of 4 nodes with 6 links used throughout this section. Red and green arrows denote negative and positive links, respectively.} \label{4nodeGraph}
	\end{center}
\end{figure}

In this appendix, we provide a step by step computation of network-level predictability for a small network. The network $G$ of 4 nodes with 6 links is given in Figure \ref{4nodeGraph}.
Its adjacency matrix $A$ is given by:
\begin{equation*}
	\begin{split}A =
		\begin{pmatrix}
			0 & 0 & 1 & 1\\
			0 & 0 & 0 & 0\\
			1 & 0 & 0 & -1\\
			-1 & -1 & 0 & 0\\
		\end{pmatrix}
	\end{split}
\end{equation*} 
Normalized left eigenvectors of $A$ are:
\begin{equation*}
	\begin{split}
		\vec e_1 &= \begin{bmatrix}
			-0.356i \\ 
			0.3062 - 0.5303i\\
			0.3062 + 0.1768i\\ 
			0.6124\\
		\end{bmatrix}, 
			\vec e_2 = \begin{bmatrix}
		0.356i \\ 
		0.3062 +0.5303i\\
		0.3062 -0.1768i\\ 
		0.6124\\
	\end{bmatrix},\\
		\vec e_3 &= \begin{bmatrix}
	-0.7071\\
	0\\
	-0.7071\\
	0\\
\end{bmatrix},  
		\vec e_4 = \begin{bmatrix}
	0\\
	1\\
	0\\
	0\\
\end{bmatrix}.
	\end{split}
\end{equation*} 
Since network size allows, Boolean dynamics is run from all possible initial conditions. State space structure is summarized in Figure \ref{4nodeGraphStateSpace}:
\begin{figure}[!htbp]
	\begin{center}
		\includegraphics[angle=-0,scale=0.35]{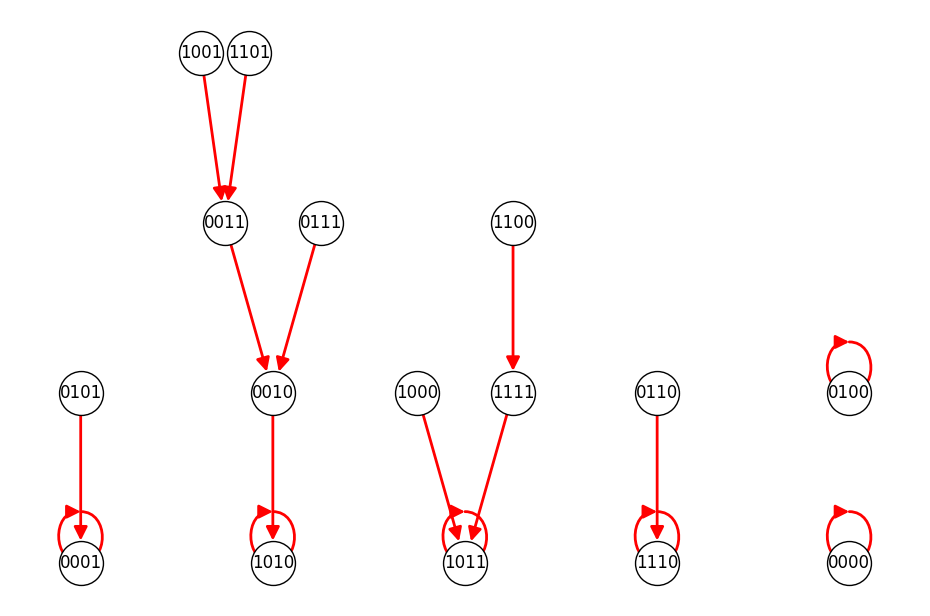}
		\caption{Structure of the state space, with nodes representing states and arrows representing transitions under the Boolean dynamical rule} \label{4nodeGraphStateSpace}
	\end{center}
\end{figure}

After discarding $\vec 0$ (for which computation of correlations is impossible), we are left with the following set of fixed-point attractors: 
\begin{equation*}
	\begin{split}
		\vec a_1 = 
		\begin{bmatrix}
			1\\ 
			0\\
			1\\ 
			0\\
		\end{bmatrix}, 
		\vec a_2 = 
		\begin{bmatrix}
			1\\ 
			1\\
			1\\ 
			0\\
		\end{bmatrix},
		\vec a_3 = 
		\begin{bmatrix}
			0\\
			0\\
			0\\
			1\\
		\end{bmatrix},  
		\vec a_4 = 
		\begin{bmatrix}
			1\\
			0\\
			1\\
			1\\
		\end{bmatrix}, 
		\vec a_5 = 
		\begin{bmatrix}
			0\\
			1\\
			0\\
			0\\
		\end{bmatrix}.
	\end{split}
\end{equation*} 
 To calculate predicability of every attractor, we compute absolute values of correlation coefficients for every pair $(\vec e_{k}, \vec a_{j})$:
 
\begin{center}
	\begin{tabular}{|c||c|c|c|c|}
		\hline
		$|\mbox{corr}(\vec e_{k}, \vec a_{j})|$& $\vec e_1$ & $\vec e_2$ & $\vec e_3$ & $\vec e_4$ \\
		\hline \hline
		$\vec a_1$ & 0.5 & 0.5 & 1 & 0.5773 \\
		\hline
		$\vec a_2$ & 0.5773 &  0.5773 & 0.5773 & 0.3333 \\
		\hline
		$\vec a_3$ & 0.5773 &  0.5773 & 0.5773 & 0.3333 \\
		\hline
		$\vec a_4$ & 0.5773 &  0.5773 & 0.5773 & 1 \\
		\hline
		$\vec a_5$ & 0.5773 &  0.5773 & 0.5773 & 1 \\
		\hline
	\end{tabular}
\end{center}
Based on the above table, $\pi(\vec a_1)= \pi(\vec a_4) = \pi(\vec a_5) = 1, \pi(\vec a_2) = \pi(\vec a_3) = 0.5773.$

In order to compute network-level predictability of $G$, predictability distribution of shuffled attractors is required. For every $a_i$, we consider all of its permutations that do not equal another $a_i$:

\begin{equation*}
	\begin{split}
		\vec a_1 &:
		\begin{bmatrix}
			1\\ 
			1\\
			0\\ 
			0\\
		\end{bmatrix}, 
		\begin{bmatrix}
			1\\ 
			0\\
			0\\ 
			1\\
		\end{bmatrix},
		\begin{bmatrix}
			0\\ 
			1\\
			1\\ 
			0\\
		\end{bmatrix},
		\begin{bmatrix}
			0\\ 
			1\\
			0\\ 
			1\\
		\end{bmatrix},
		\begin{bmatrix}
			0\\ 
			0\\
			1\\ 
			1\\
		\end{bmatrix}, \\    
		\vec a_2 &: 
		\begin{bmatrix}
			1\\ 
			1\\
			0\\ 
			1\\
		\end{bmatrix},
		\begin{bmatrix}
			0\\ 
			1\\
			1\\ 
			1\\
		\end{bmatrix},\:\:\:
		\vec a_3: 
		\begin{bmatrix}
			1\\
			0\\
			0\\
			0\\
		\end{bmatrix},
		\begin{bmatrix}
			0\\ 
			0\\
			1\\ 
			0\\
		\end{bmatrix}, \\ 
		\vec a_4 &:
		\begin{bmatrix}
			1\\ 
			1\\
			0\\ 
			1\\
		\end{bmatrix},
		\begin{bmatrix}
			0\\ 
			1\\
			1\\ 
			1\\
		\end{bmatrix},\:\:\: 
		\vec a_5:
		\begin{bmatrix}
			1\\
			0\\
			0\\
			0\\
		\end{bmatrix},
		\begin{bmatrix}
			0\\ 
			0\\
			1\\ 
			0\\
		\end{bmatrix}.
	\end{split}
\end{equation*} 
Distribution of $\pi(\vec a)$ for those shuffled attractors has $\langle C \rangle = 0.6543$ and $\sigma = 0.1437$. Using the $2\sigma$ threshold yields network-level predictability $\pi(G) = 0.6.$

\section{Example of individual positive and negative cycles}

\begin{figure}[!htbp]
	\begin{center}
		\subfloat[b][]{
			\includegraphics[angle=-0,scale=0.3]{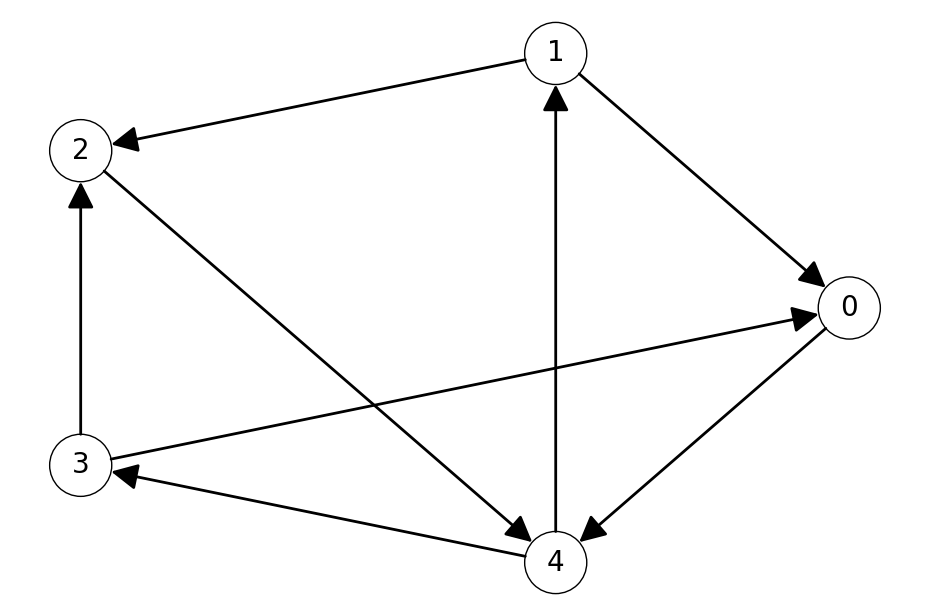}
		}\\
		\subfloat[b][]{
			\includegraphics[angle=-0,scale=0.3]{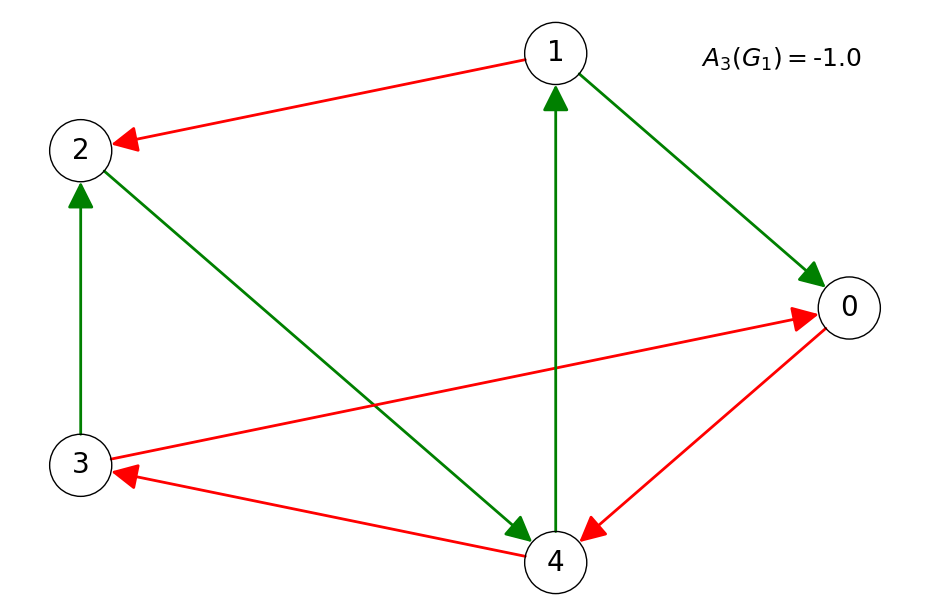}
		}\\
		\subfloat[b][]{
				\includegraphics[angle=-0,scale=0.3]{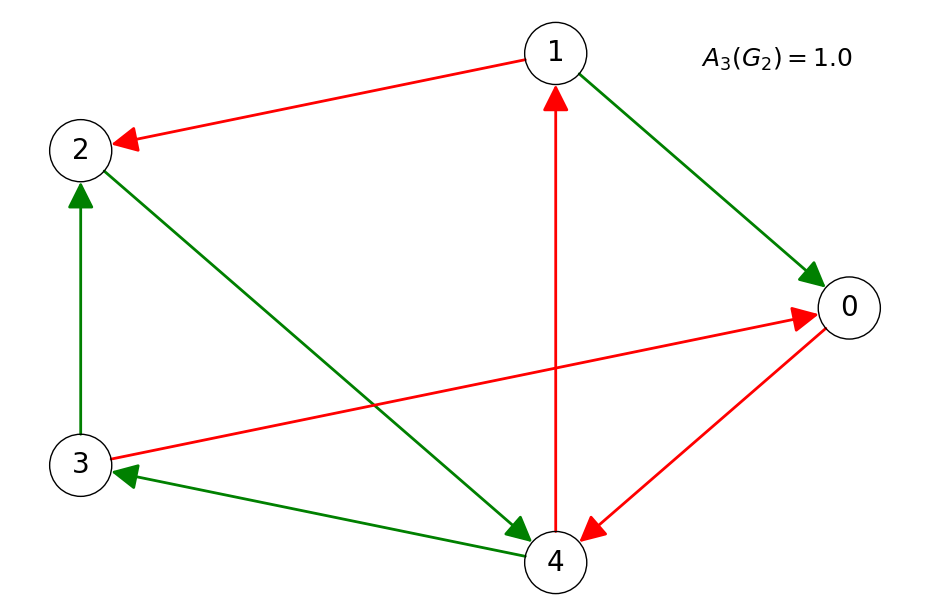}
		}
		\caption{(A) Directed topology of a network on 5 nodes with 8 links to used in this section. (B) Arrangement of link signs $G_1$ with 3-cycle asymmetry $A_3(G_1)=-1$. Red and green arrows denote negative and positive links, respectively. (C) Arrangement of link signs $G_2$ with 3-cycle asymmetry $A_3(G_2)=1$. Red and green arrows denote negative and positive links, respectively.}\label{signVariants}
	\end{center}
\end{figure}

In this appendix, we illustrate for small graphs how alteration of 3-cycle asymmetry changes network-level predictability of a small network. To that end, we fix the directed unsigned topology depicted in the Figure \ref{signVariants}A, and endow it with different arrangements of links signs. 

For the sake of consistency with the main text, we only consider sign arrangements which have equal number of positive and negative links, and refer to those as \textit{balanced} sign arrangement. 

\begin{figure}[!htbp]
	\begin{center}
		\includegraphics[angle=-0,scale=0.3]{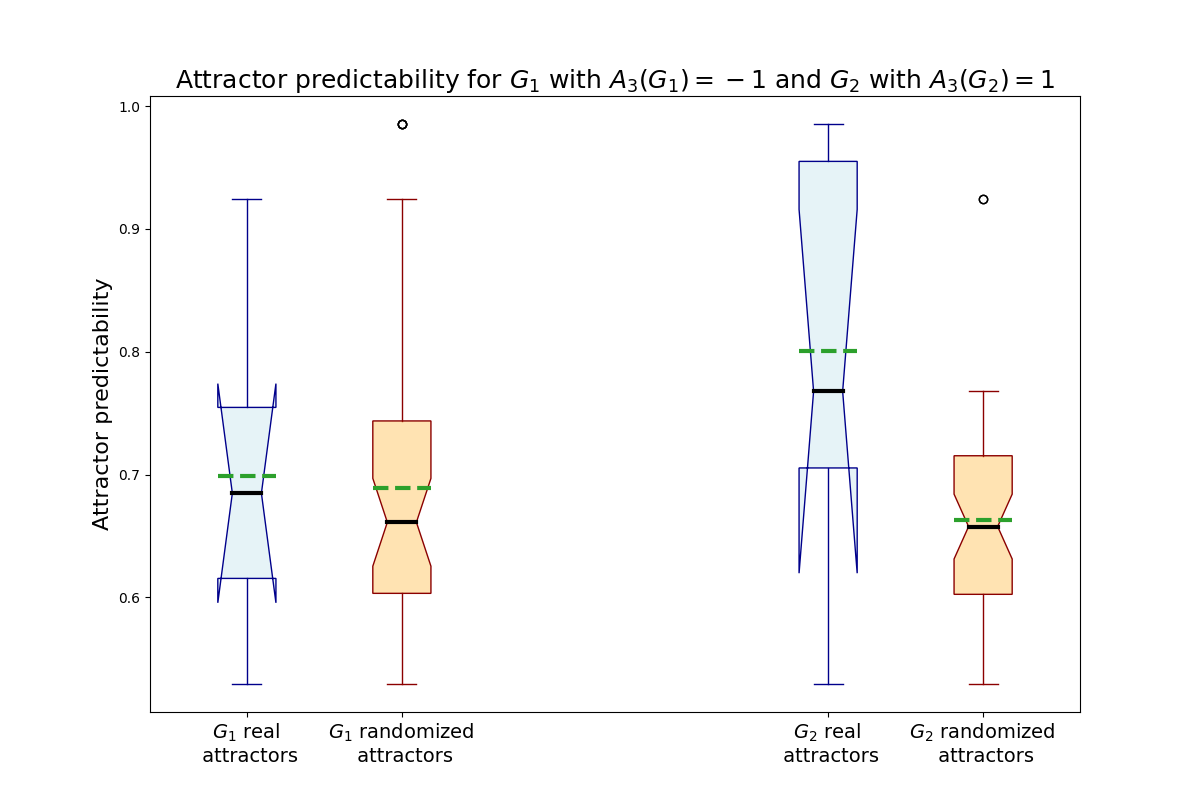}
		\caption{Predictability of actual and shuffled attractors for $G_1$ and $G_2$, computed similarly to Appendix A. Green and black lines represent means and medians, respectively.} \label{asymPredAppendix}
	\end{center}
\end{figure}

Figure \ref{signVariants}B features an example of a balanced sign arrangement $G_1$ where $A_3(G_1)=-1$, i.e. every 3-cycle is negative. Predictability of observed and shuffled attractors is shown in Figure \ref{asymPredAppendix}. Resulting network-level predictability $\pi(G_1)=0$

Conversely, if we consider a sign arrangement $G_2$ with $A_3(G_2)=1$ (see Figures \ref{signVariants}C, \ref{asymPredAppendix}), we see a significant increase in network-level predictability ($\pi(G_2) = 0.42$).

\begin{figure}[!hbtp]
	\begin{center}
		\includegraphics[angle=-0,scale=0.3]{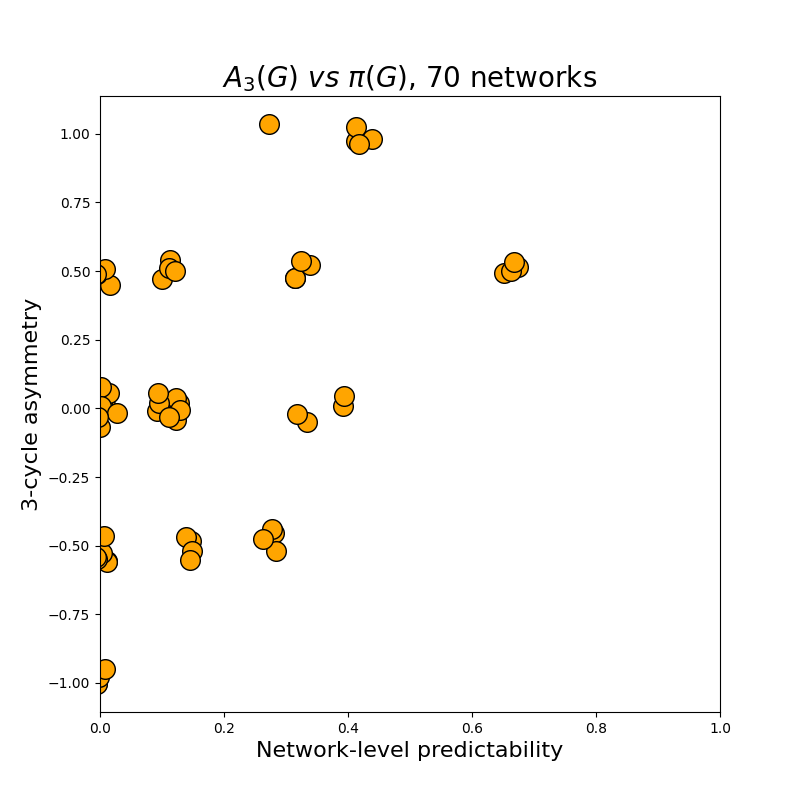}
		\caption{Scatterplot of 3-cycle asymmetry and network-level predictability of all 70 labeled balanced sign arrangements of topology in Figure \ref{signVariants}A. To clearly show the number of networks with coinciding pairs of values, small random shifts have been applied. } \label{balancedScatter}
	\end{center}
\end{figure}

To verify existence of relationship between $A_3(G)$ and $\pi(G)$, we construct all labeled balanced link sign arrangements on the directed topology shown in Figure \ref{signVariants}A.  The scatterplot of 3-cycle asymmetry and network-level predictability for those arrangements is featured in Figure \ref{balancedScatter}. Based on cycle asymmetry, two distinct groups of networks can be naturally defined: one with $A_3(G)<=-0.5$ and another with $A_3(G)>=0.5$. Between those two groups, distributions of network-level predictability are markedly different: networks from the first group dominantly exhibit low values of $\pi(G)$, while networks of the second tend to have high network-level predictability. Pearson correlation of $A_3(G)$ and $\pi(G)$ is equal to $0.504$ if all balanced networks are considered, and becomes $0.4618$ without duplicate value pairs.

\clearpage
\bibliography{boolean_predictability}


\setcounter{figure}{0}
\renewcommand{\thefigure}{S\arabic{figure}}
\setcounter{table}{0}
\renewcommand{\thetable}{S\arabic{table}}

\end{document}


\title{Predicting attractors from spectral properties of stylized gene regulatory networks\protect\\  Supplemental material}
	
	\author{Dzmitry Rumiantsau}
	\email{d.rumiantsau@constructor.university}
	\affiliation{%
		Department of Life Sciences and Chemistry, Constructor University, D-28759 Bremen, Germany}
	\author{ Annick Lesne}%
	\email{annick.lesne@sorbonne-universite.fr}
	\affiliation{%
		Sorbonne Universit\'e, CNRS,  Laboratoire de Physique Th\'eorique de la Mati\`ere Condens\'ee, LPTMC, F-75252, Paris, France}%
	\affiliation{%
		Institut de G\'en\'etique Mol\'eculaire de Montpellier, University of Montpellier, CNRS, F-34293, Montpellier, France\looseness=-1}%
	
	\author{Marc-Thorsten H\"utt}
	\email{m.huett@constructor.university}
	\affiliation{Department of Life Sciences and Chemistry, Constructor University, D-28759 Bremen, Germany}
	
	\date{\today}
	\maketitle
	
	
	\noindent

	\begin{center}
				\begin{longtable}{|c|c|c|c|c|}
						\caption{Network of 40 nodes with 80 positive and 80 negative directed links used to illustrate the definition of network-level predictability in Figure 2B. The first cell in each row features the node number. The next four cells contain lists of other endpoints of positive outgoing, negative outgoing, positive incoming and negative incoming links for the node, respectively.}\label{supTable1}\\
			\hline
			\textbf{Node} & \textbf{Out +} & \textbf{Out -} & \textbf{In +} & \textbf{In -}\\
			\hline
			0 & 27,33 & 14,25,34,35 &  & 10,16,20,21,32,33,34
\\ 
			\hline 
			1 & 12,14,25 & 37 & 10 & 14
\\ 
			\hline 
			2 &  & 12 & 5,23 & 14
\\ 
			\hline 
			3 & 18,21,34,36 & 23,29,30 &  & 14
\\ 
			\hline 
			4 & 20,28,31,33 & 5 &  & 25
\\ 
			\hline 
			5 & 2,13,23,31,38 & 20,25 & 7,25,27,35 & 4,26
\\ 
			\hline 
			6 & 24,32,35,37 & 8,12,26,38 &  & 
\\ 
			\hline 
			7 & 5,12 & 13,14,24,25,31 & 9,27,39 & 
\\ 
			\hline 
			8 & 37 & 18,33 & 9 & 6,23,24,31
\\ 
			\hline 
			9 & 7,8 & 14 & 15,27 & 10
\\ 
			\hline 
			10 & 1,19,33 & 0,9,13,32,36 & 18 & 17,24,26,29
\\ 
			\hline 
			11 & 14,17,38 &  &  & 15,37
\\ 
			\hline 
			12 & 31 & 17 & 1,7 & 2,6,16,32
\\ 
			\hline 
			13 & 20,30,34 & 29 & 5,21 & 7,10
\\ 
			\hline 
			14 &  & 1,2,3,15,25 & 1,11,27 & 0,7,9,33
\\ 
			\hline 
			15 & 9,25 & 11,26 & 16,35 & 14
\\ 
			\hline 
			16 & 15,29 & 0,12,36,39 & 32 & 27,33
\\ 
			\hline 
			17 & 23,27,37 & 10 & 11 & 12,27
\\ 
			\hline 
			18 & 10,19,28,39 &  & 3 & 8,33
\\ 
			\hline 
			19 & 24,35 &  & 10,18 & 26
\\ 
			\hline 
			20 & 27,33 & 0,23,28 & 4,13 & 5,21,28
\\ 
			\hline 
			21 & 13,28 & 0,20,33,39 & 3 & 28
\\ 
			\hline 
			22 & 38 &  &  & 25
\\ 
			\hline 
			23 & 2 & 8,28 & 5,17,36 & 3,20
\\ 
			\hline 
			24 & 39 & 8,10 & 6,19,25 & 7
\\ 
			\hline 
			25 & 5,24,35,38 & 4,22 & 1,15,35,37 & 0,5,7,14
\\ 
			\hline 
			26 &  & 5,10,19 &  & 6,15
\\ 
			\hline 
			27 & 5,7,9,14,29 & 16,17,32 & 0,17,20,29 & 
\\ 
			\hline 
			28 &  & 20,21 & 4,18,21,32 & 20,23
\\ 
			\hline 
			29 & 27 & 10,36 & 16,27,36 & 3,13,30
\\ 
			\hline 
			30 &  & 29 & 13 & 3,39
\\ 
			\hline 
			31 &  & 8 & 4,5,12 & 7,35,36
\\ 
			\hline 
			32 & 16,28 & 0,12 & 6 & 10,27
\\ 
			\hline 
			33 &  & 0,14,16,18 & 0,4,10,20,37 & 8,21
\\ 
			\hline 
			34 &  & 0,35 & 3,13 & 0
\\ 
			\hline 
			35 & 5,15,25 & 31 & 6,19,25 & 0,34
\\ 
			\hline 
			36 & 23,29 & 31 & 3,37,39 & 10,16,29
\\ 
			\hline 
			37 & 25,33,36 & 11 & 6,8,17,39 & 1
\\ 
			\hline 
			38 &  &  & 5,11,22,25 & 6
\\ 
			\hline 
			39 & 7,36,37 & 30 & 18,24 & 16,21
\\ 
			\hline 
			
		\end{longtable}
	\end{center}

%

	\begin{figure*}[hbtp]
		\begin{center}
					\includegraphics[angle=-0,scale=0.5]{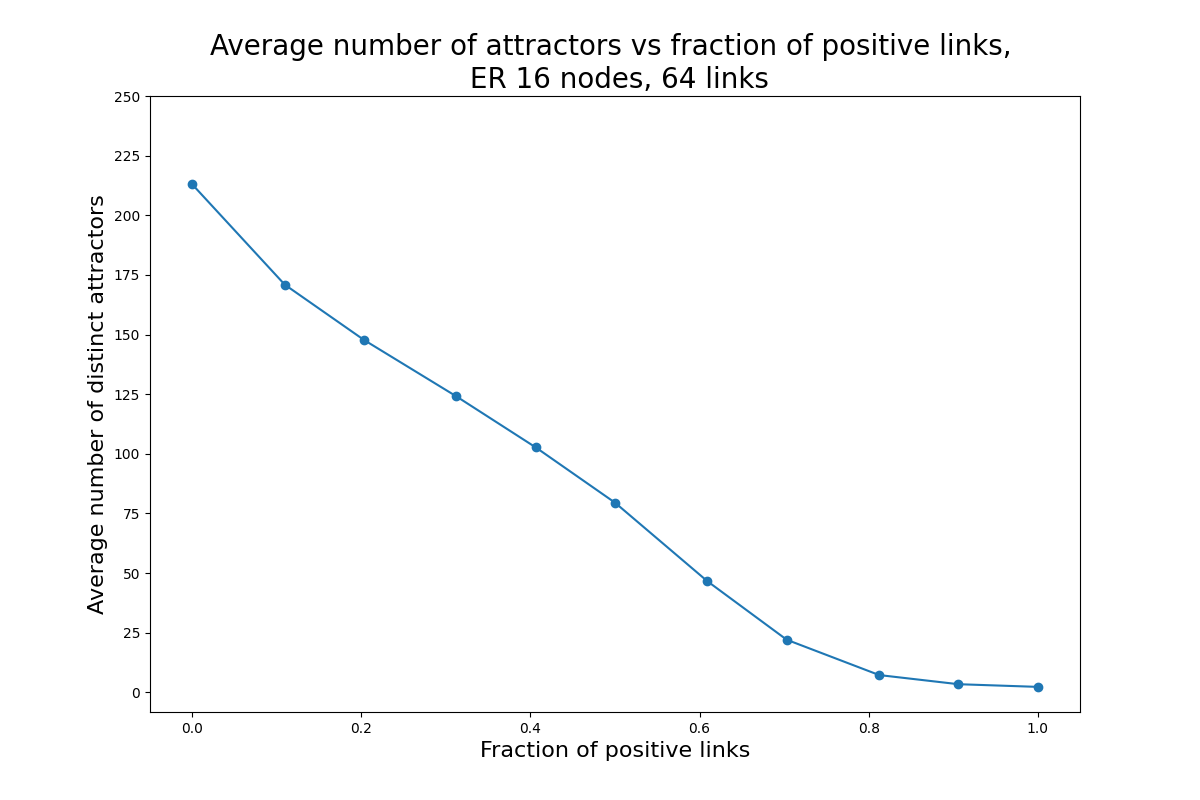}
			\caption{Average number of distinct fixed point attractors of boolean dynamics from \cite{li2004yeast} depending on the proportion of positive links in the network $p_+$. For each value of $p_+$ between 0 and 1 with $0.1$ increments, 10.000 random ER networks with 16 nodes and 64 links have been considered.} \label{supFigure1}
		\end{center}
	\end{figure*}
	\begin{figure*}[!hbtp]
	\begin{center}
		\includegraphics[angle=-0,scale=0.5]{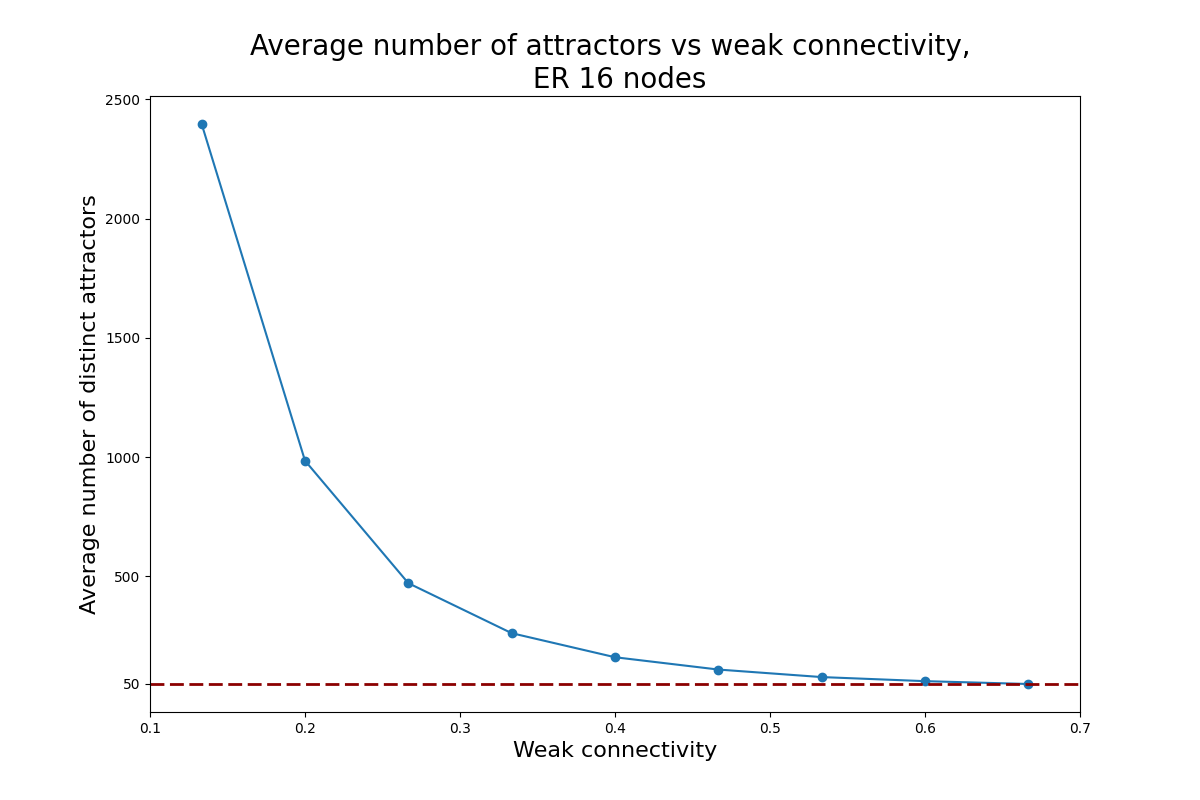}
		\caption{Average number of distinct fixed point attractors under \cite{li2004yeast} depending on network connectivity. For each number of links between 16 and 80 with increments by 8, 10.000 random ER networks with 16 nodes and this number of links and equal number of positive and negative links have been considered.} \label{supFigure2}
	\end{center}
	\end{figure*}
	\begin{figure*}[!hbtp]
	\begin{center}
		\includegraphics[angle=-0,scale=0.5]{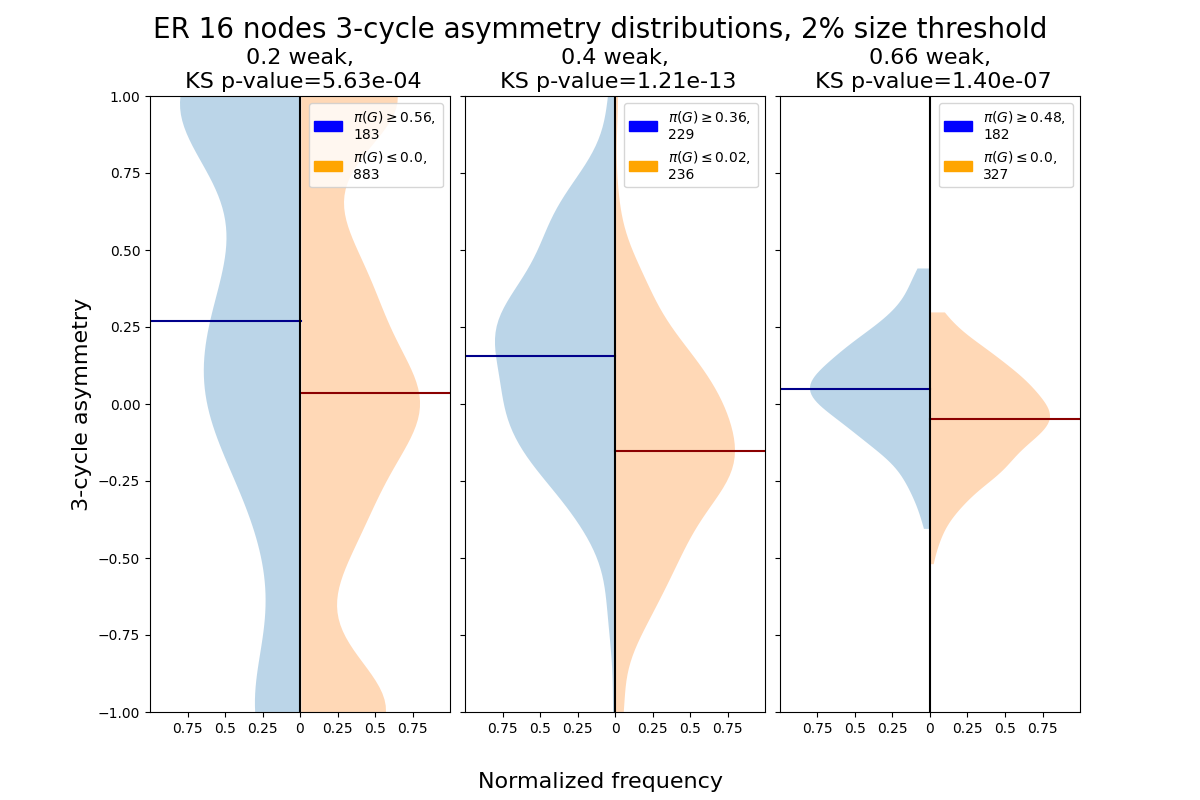}
		\caption{Distributions of 3-cycle asymmetry for sets of networks with high and low network-level predictability and different connectivity values. For 26, 48 and 80 links each (0.2, 0.4 and 0.66 weak connectivity), 10.000 random ER networks with 16 nodes and that many links have been considered. Minimum numbers of distinct attractors required to enter the histogram are 250, 50 and 20, respectively. Numbers of positive and negative links are equal. KS p-value corresponds to the p-value obtained from two-sample Kolmogorov–Smirnov test. We define sets of networks with high and low network-level predictability by using the 2\% quantile-motivated thresholds (see Figure 2 in the main text).} \label{supFigure3}
	\end{center}
    \end{figure*}
	\begin{figure*}[!hbtp]
	\begin{center}
		\includegraphics[angle=-0,scale=0.5]{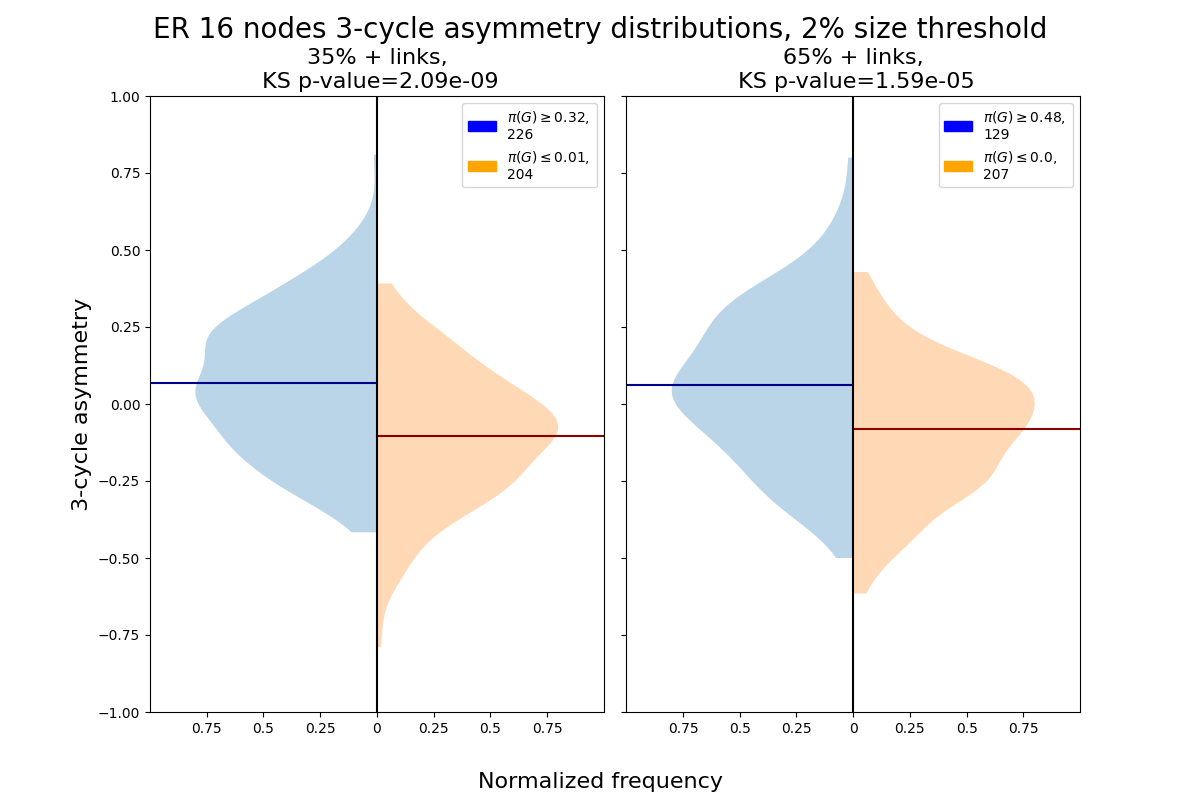}
		\caption{Distributions of 3-cycle asymmetry for sets of networks with high and low network-level predictability and different fraction of positive links $p_+$. 10.000 random ER networks with 16 nodes and 64 links with 35\% positive links and 10.000 networks with 16 nodes and 64 links with $p_+=0.65$ have been considered. KS p-value corresponds to the p-value obtained from two-sample Kolmogorov–Smirnov test. We define sets of networks with high and low network-level predictability by using the 2\% quantile-motivated thresholds (see Figure 2 in the main text).} \label{supFigure4}
	\end{center}
	\end{figure*}
	\begin{figure*}[!hbtp]
	\begin{center}
		\includegraphics[angle=-0,scale=0.5]{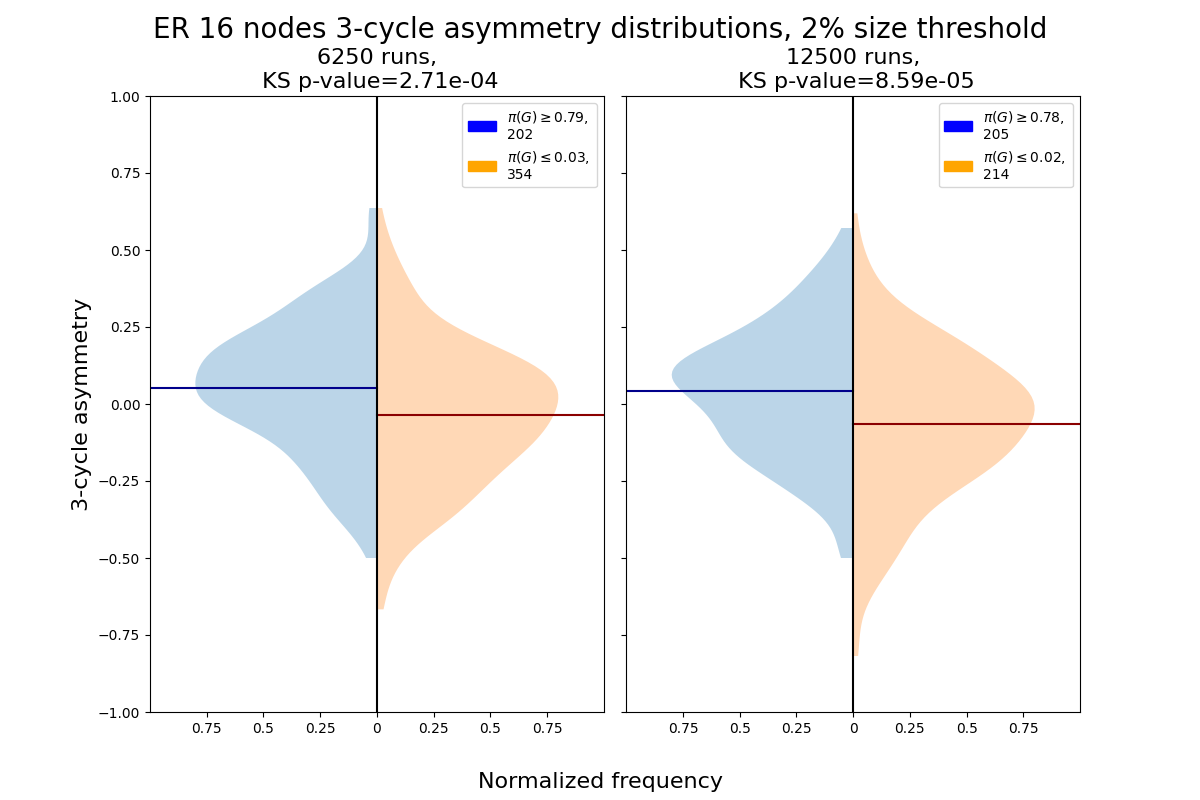}
		\caption{Distributions of 3-cycle asymmetry for sets of networks with high and low network-level predictability and different number of initial states. For the same set of 10.000 random ER networks with 40 nodes and 160 links with $p_+=0.5$ calculation of network-level predictability is performed two distinct times - using 6250 and 12500 random initial states. KS p-value corresponds to the p-value obtained from two-sample Kolmogorov–Smirnov test. We define sets of networks with high and low network-level predictability by using the 2\% quantile-motivated thresholds (see Figure 2 in the main text).} \label{supFigure5}
	\end{center}
	\end{figure*}
	\begin{figure*}[!hbtp]
	\begin{center}
		\includegraphics[angle=-0,scale=0.5]{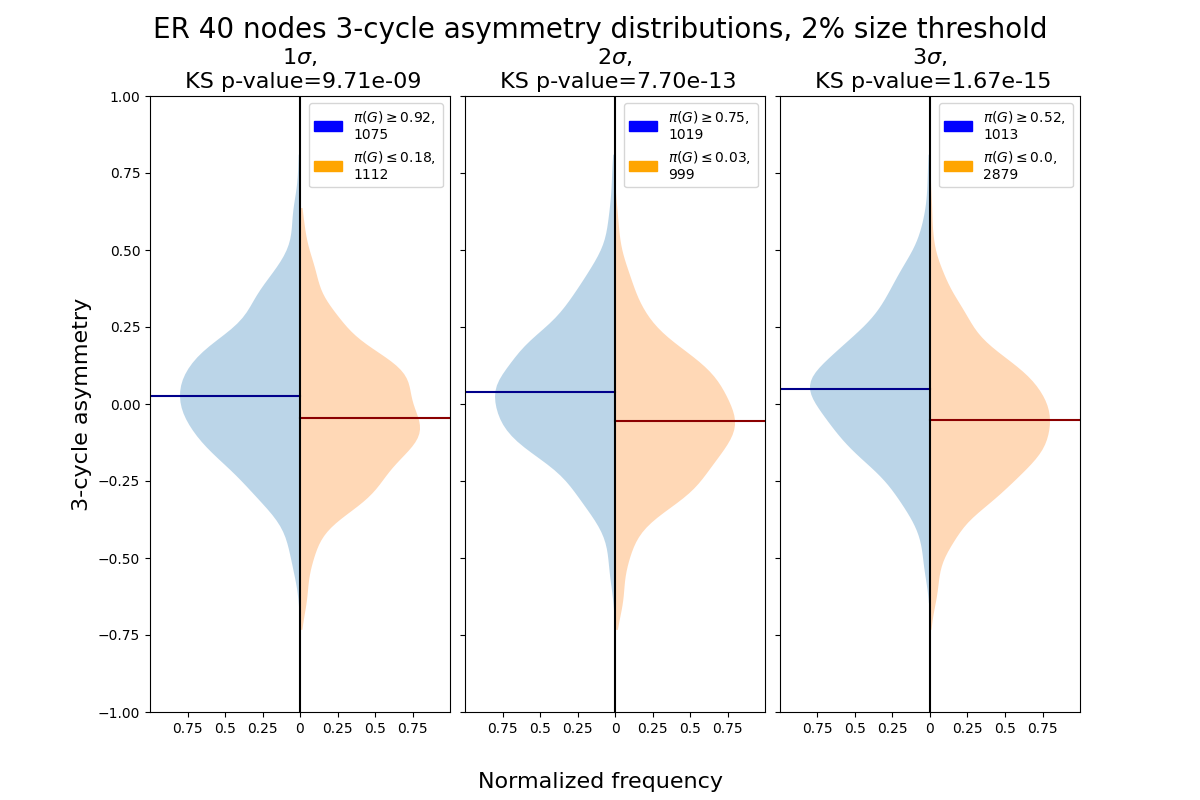}
		\caption{Distributions of 3-cycle asymmetry for sets of networks with high and low network-level predictability and random sampling of initial conditions. A set of 50.000 ER networks on 40 nodes with 80 positive and 80 negative links has been considered. For each network dynamics was run starting from 50.000 random initial states. Minimal number of observed distinct attractors required to enter this figure is 100. Network-level predictability is displayed for thresholds of 1, 2 and 3 standard deviations away from randomness. KS p-value corresponds to the p-value obtained from two-sample Kolmogorov–Smirnov test. We define sets of networks with high and low network-level predictability by using the 2\% quantile-motivated thresholds (see Figure 2 in the main text).} \label{supFigure6}
	\end{center}
	\end{figure*}
\begin{figure}[!htbp]
	\begin{center}
		\includegraphics[angle=-0,scale=0.5]{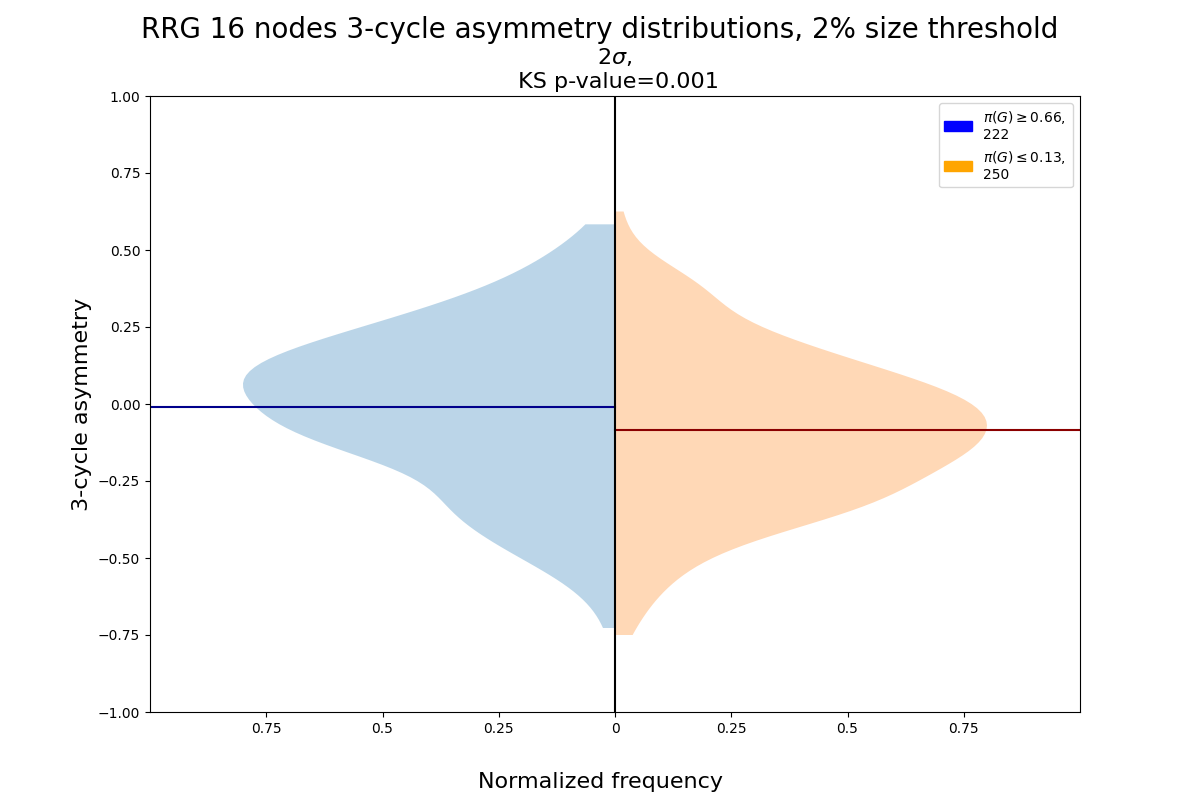}
		\caption{Distributions of 3-cycle asymmetry for sets of random regular networks with high and low network-level predictability values. 10.000 signed directed regular random networks with 16 nodes and degree 2 (64 links total) have been considered. We define sets of networks with high and low network-level predictability by using the 2\% quantile-motivated thresholds (see Figure 2 in the main text).} \label{supFigure9}
\end{center}\end{figure}
	\begin{figure*}[!hbtp]
	\begin{center}
		\includegraphics[angle=-0,scale=0.7]{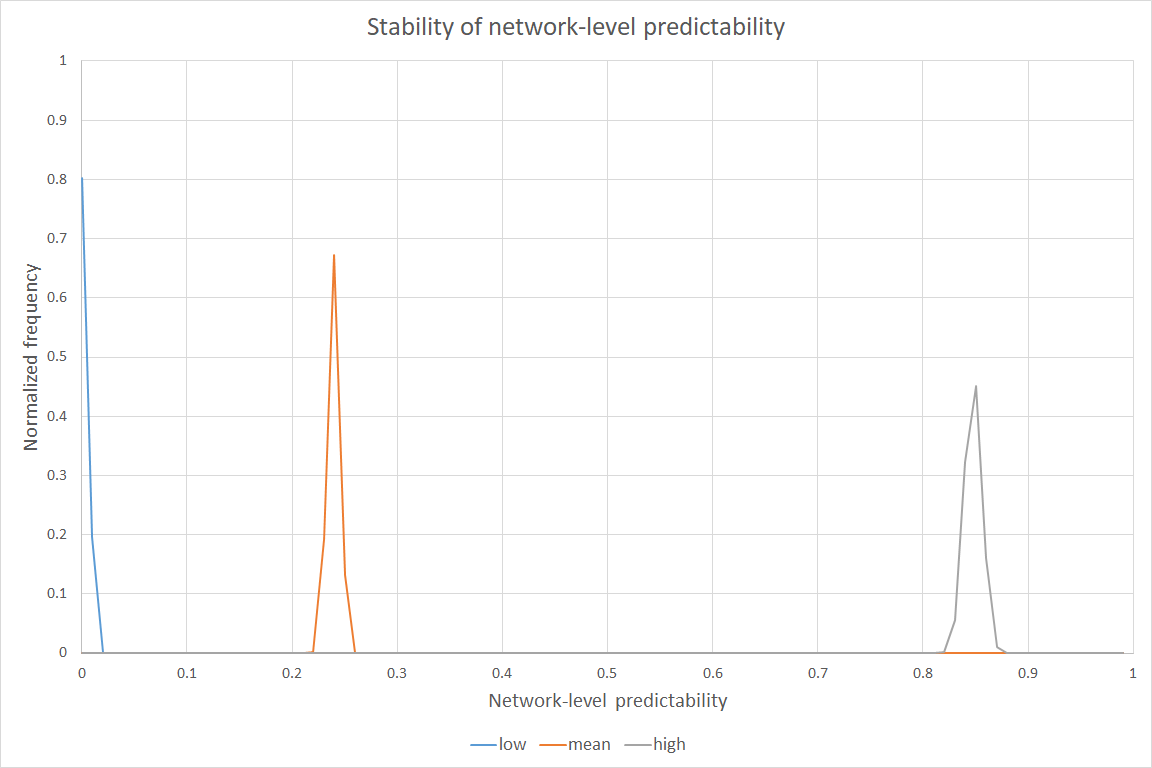}
		\caption{Stability of network-level predictability. Based on the distribution of network-level predictability for ER networks with 40 nodes and 80 positive and 80 negative links, we chose 3 sample networks: one from the mean and one from each extreme set. Afterwards, 10.000 different sets of network-level predictability computations with independently chosen sets of 50.000 initial conditions have been performed for each of those networks. This figure shows resulting distributions of network-level predictabilities, sharply centered around original values.}\label{supFigure7}
	\end{center}
	\end{figure*}
	\begin{figure*}[!hbtp]
	\begin{center}
		\includegraphics[angle=-0,scale=0.7]{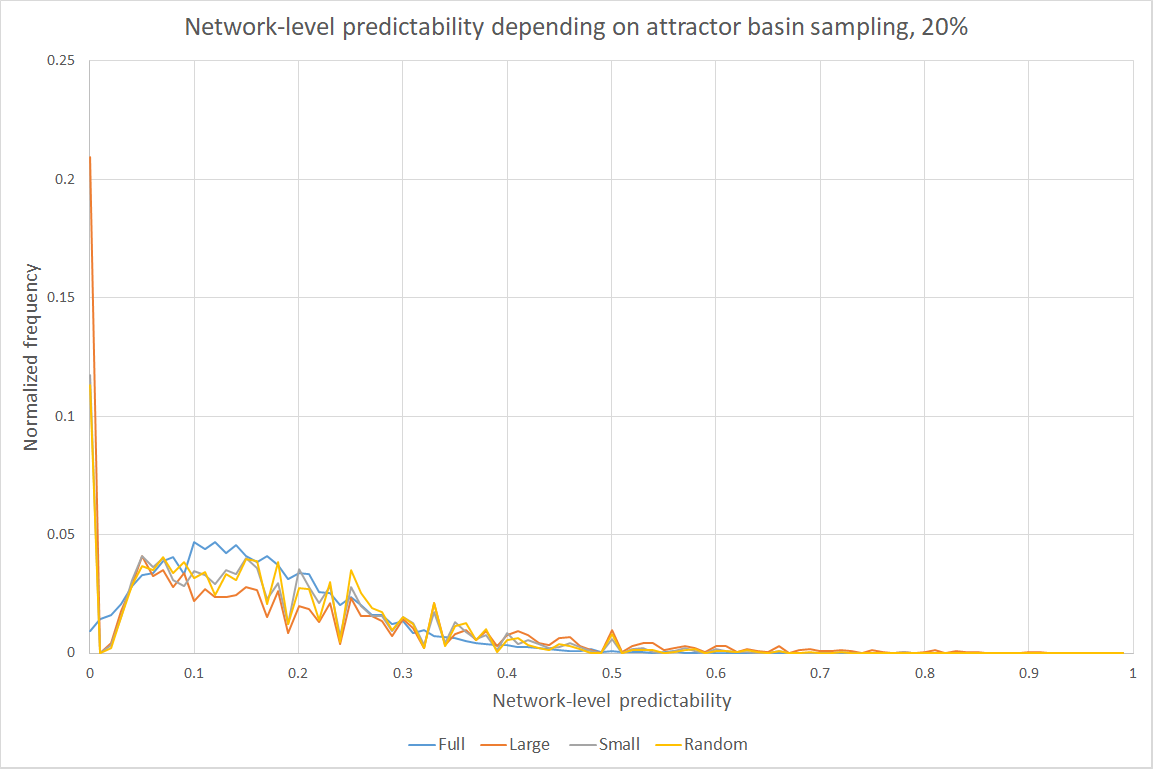}
		\caption{Influence of basin size on network-level predictability. Blue curve shows distribution of $\pi(G)$ for 10.000 networks with 16 nodes and 32 positive and 32 negative links and exhaustive sampling of initial conditions using predictabilities of all attractors. Other curves show distributions of $\pi(G)$ calculated using 20\% of attractors with largest, smallest or randomly chosen basins for the same set of networks. It can be seen that using only large-basin steady states somewhat amplifies extremes of the distribution compared to other strategies, but all three samplings have the same qualitative properties. }\label{supFigure8}
	\end{center}
	\end{figure*}

\clearpage
	\textbf{GeneNetWeaver model descriptions}

The original model simulates gene regulation. For each gene $i$ of a network, the rate of change of mRNA concentration $ F_{i}^{RNA} $ and the rate of change of protein concentration $F_{i}^{Prot}$ are described by: 
\begin{equation*}
	\begin{split}
		F_{i}^{RNA} &= \dfrac{\mathrm{d}x_{i}}{\mathrm{d}t}=m_{i}\cdot f_{i}(y) - \lambda_{i}^{RNA}\cdot x_{i}\\
		F_{i}^{Prot} &= \dfrac{\mathrm{d}y_{i}}{\mathrm{d}t}=r_{i}\cdot x_{i} - \lambda_{i}^{Prot}\cdot y_{i},
	\end{split}
\end{equation*}
where $ m_{i} $ is the maximum transcription rate, $ r_{i} $ the translation rate, $\lambda_{i}^{RNA}$ and $ \lambda_{i}^{Prot} $ are the mRNA and protein degradation rates and $ x $ and $ y $ are vectors containing all mRNA and protein concentration levels, respectively. $f_{i}(\cdot)$ is the activation function of a gene, which computes the relative activation of a gene, which is between 0 (the gene is shut off) and 1 (the gene is maximally activated). This system of equations is then integrated to obtain noiseless mRNA and protein concentration levels for each gene \cite{schaffter2011genenetweaver}. In the original paper, the model is further stochastically enhanced by simulating molecular noise, but this enhancement is not applied in the current study. 

The above model was selected because there is a freely available tool \textit{Gene Net Weaver} (GNW) \cite{schaffter2011genenetweaver} that uses it to run simulations. In Gene Net Weaver, the activation function of a gene $f_{i}$ is implemented as follows: 
\begin{equation}
	\begin{split}
		f_{i}(y)=\dfrac{\prod_{j=1}^{\alpha_i}\xi (k^i_j)}{\prod_{j=1}^{n_i}(\xi (k^i_j) + 1)},
	\end{split}
\end{equation}
where $i$ has $n_i$ regulators $k^i_j$, and $\xi (k^i_j)$ are their concentrations. The first $\alpha_i$ regulators are considered to be activators, the rest are inhibitors \cite{schaffter2011genenetweaver}. While this approach preserves excitatory or inhibitory nature of gene regulation,  zero concentration of a single activator will result in deactivation of the regulated gene independently of all other regulators. This is contrary to the nature of Boolean threshold dynamics we discuss in the paper, where the sum of the regulator states, rather than their product, determines the new state of the gene. Two models can be brought closer in the way described below.

For each regulator $k^i_j$  of gene $i$ we define an \textit{elementary activation function}:
\begin{equation*}
	\begin{split}
		g_{i}(j)=\begin{cases}
			\dfrac{\xi(k^i_j)}{\xi(k^i_j) + 1}, &\text{ $k^i_j$ is an activator of $i$}\\
			\dfrac{1}{\xi(k^i_j) + 1}, &  \text{$k^i_j$ is an inhibitor of $i$}.
		\end{cases}
	\end{split}
\end{equation*}
The original activation function from (2) can be viewed as a product of all elementary activation functions over all regulators. To make the ODE model closer to the Boolean one, the product may be replaced by arithmetic mean: 
\begin{equation*}
	\begin{split}
		f_i(y) = \dfrac{\sum_{j=1}^{m_i}g_i(j)}{n_i}
	\end{split}
\end{equation*}
In the paper, the modified ODE model is referred to as the "averaging ODE model".
	\begin{figure*}[!hbtp]
	\begin{center}
		\includegraphics[angle=-0,scale=0.55]{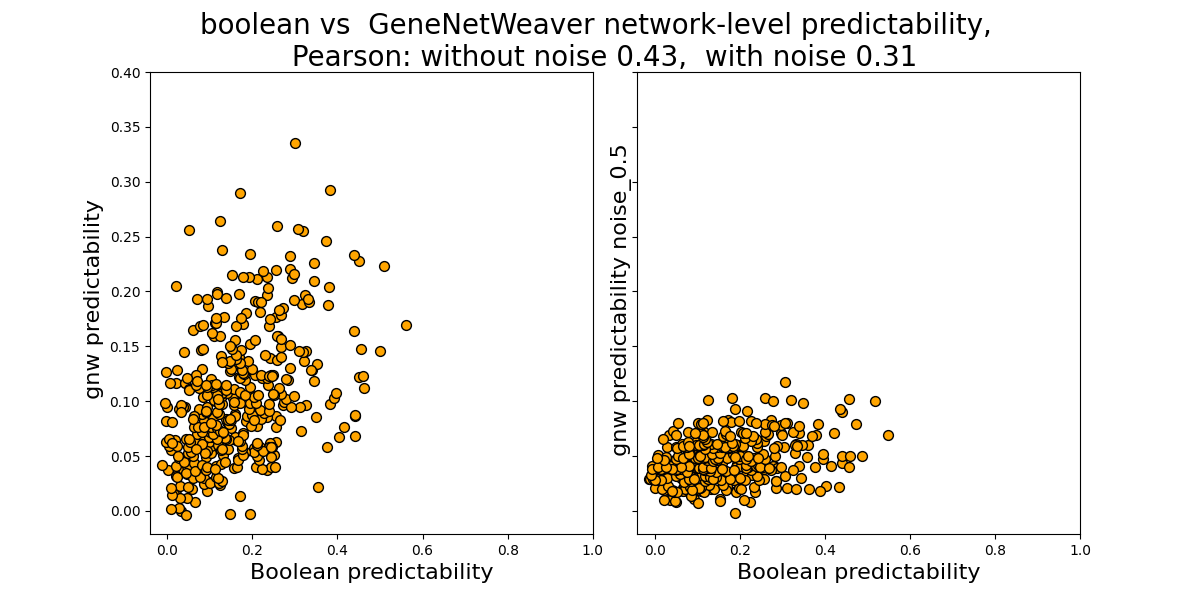}
\caption{Comparison of network-level attractor predictabilities derived from the Boolean model with those derived from the GeneNetWeaver data simulator (original model). (A) GeneNetWeaver simulation with measurement noise but no intrinsic noise. (B) GeneNetWeaver simulation with both measurement noise and intrinsic noise. Default values are used for measurement noise, while intrinsic noise corresponds to noiseCoefficientSDE=0.5 in GeneNetWeaver settings (10 times the default value)}\label{supFigure10}
	\end{center}
\end{figure*}
\begin{figure*}[!htbp]
	\begin{center}
		\subfloat[][]{
			\includegraphics[angle=-0,scale=0.25]{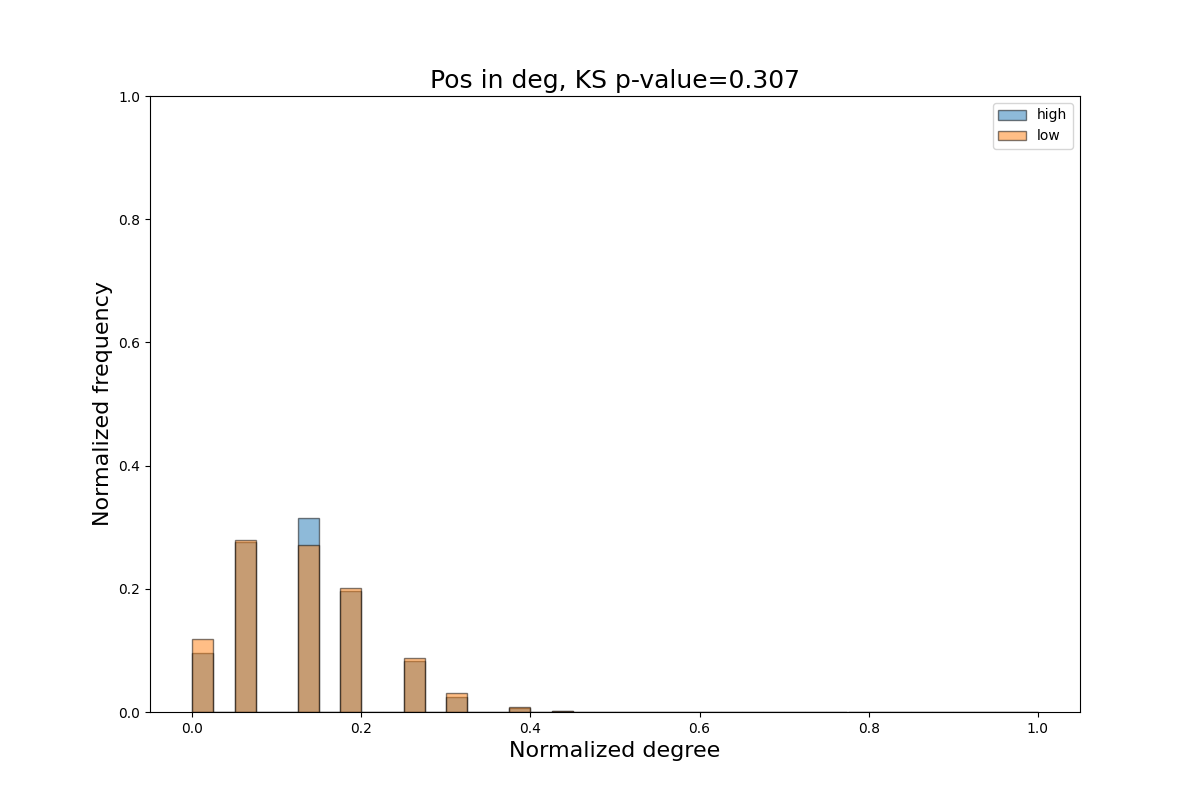}
		}
		\subfloat[][]{
			\includegraphics[angle=-0,scale=0.25]{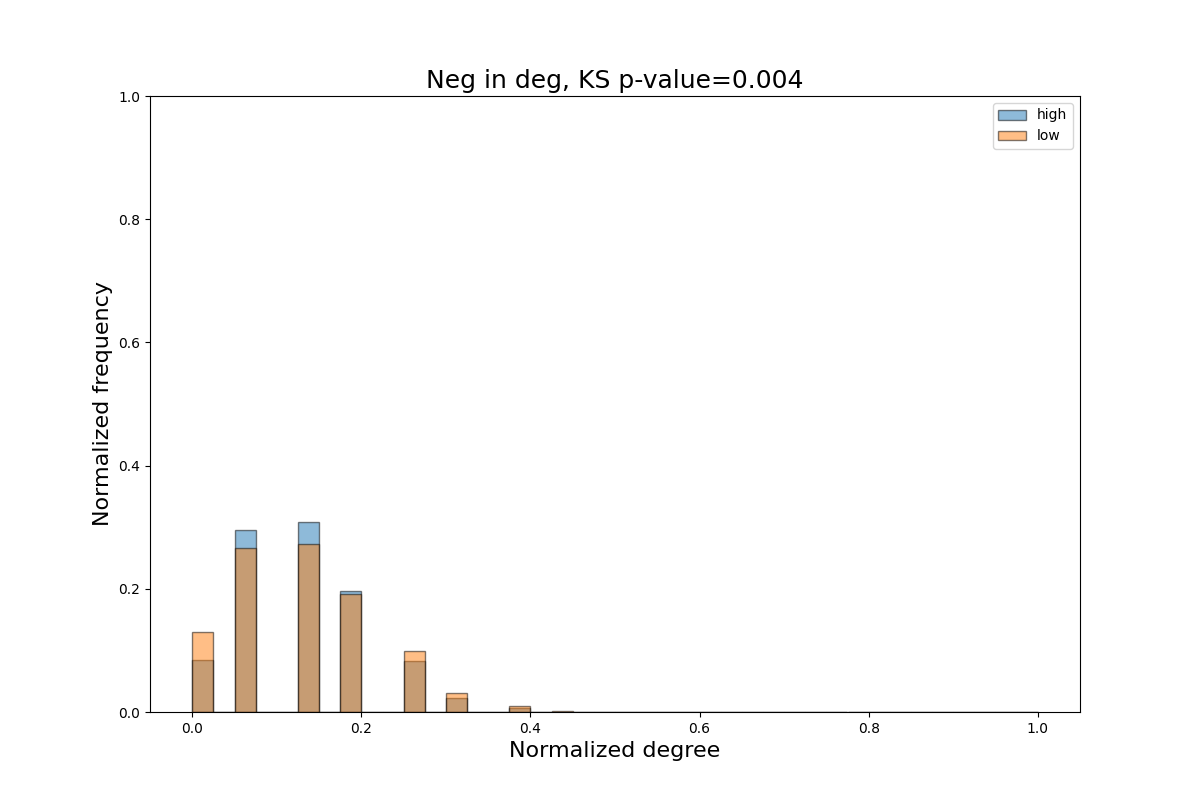}
		}\\
		\subfloat[][]{
			\includegraphics[angle=-0,scale=0.5]{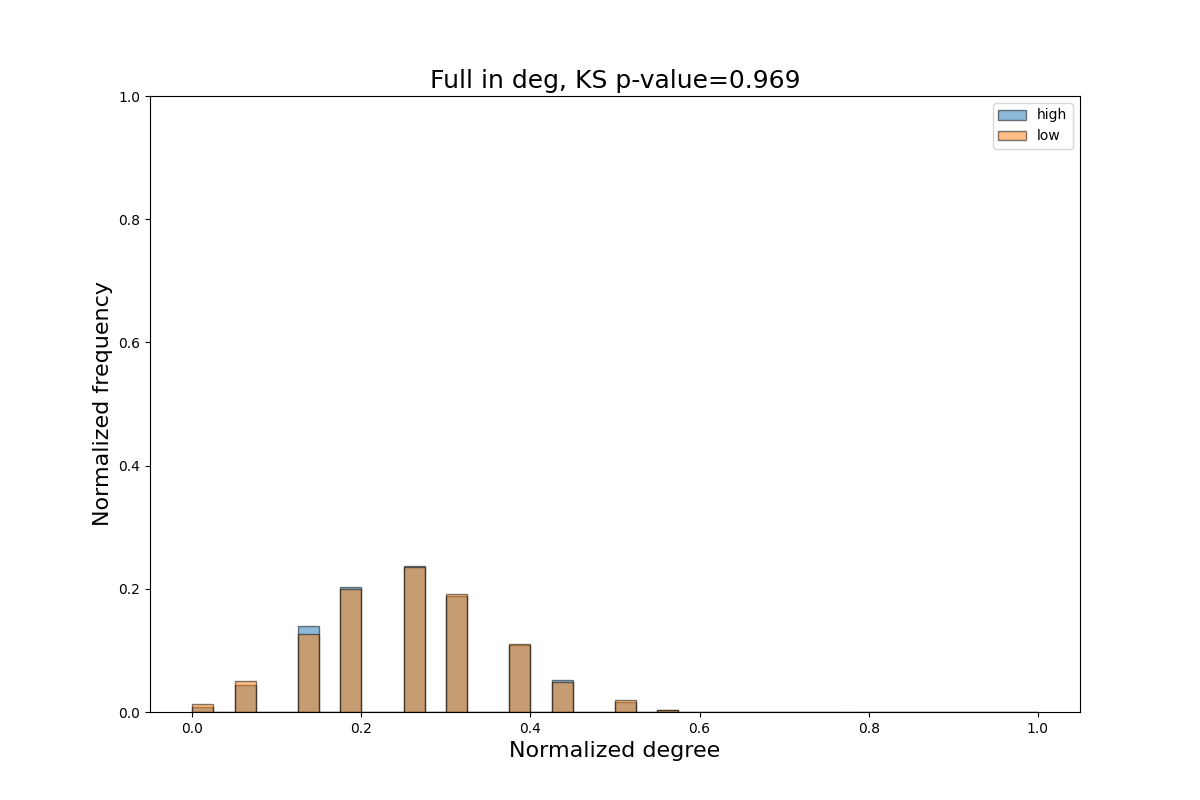}
		}
			\caption{Distributions of positive, negative and full in-degree for sets of networks with high and low network-level predictability (2\% size threshold was used for defining the set). 10.000 random ER networks with 16 nodes, 32 positive and 32 negative links were considered. Although negative in-degree distributions for those sets are different with significance below 0.01 threshold, we do not believe this quantity adequately explains the drastic difference in observed network-level predictabilities (and it is meaningless in case of signed directed regular random graphs).} \label{supFigure11}
		\end{center}
	\end{figure*}
	\begin{figure*}[!hbtp]
	\begin{center}
		\includegraphics[angle=-0,scale=0.55]{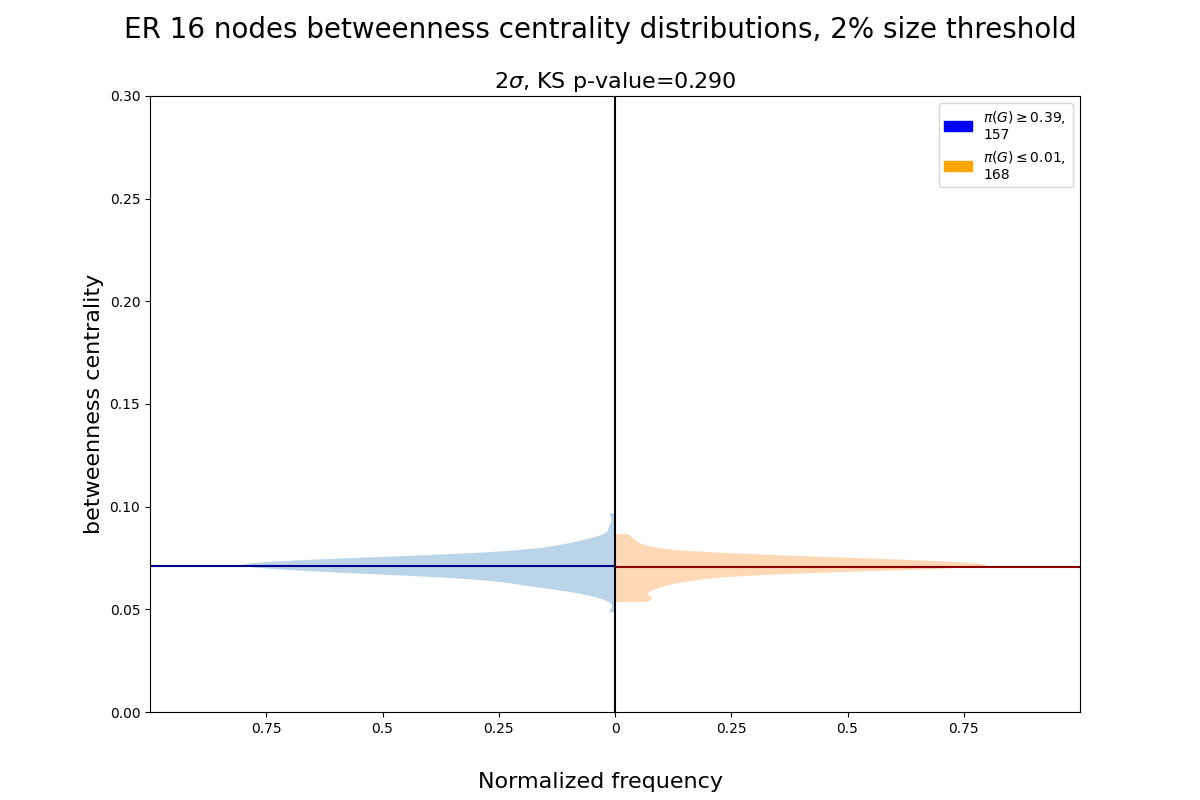}
		\caption{Distributions of average betweenness centrality for sets of networks with high and low network-level predictability. 10.000 random ER networks with 16 nodes, 32 positive and 32 negative links were considered. KS p-value corresponds to the p-value obtained from two-sample Kolmogorov–Smirnov test. We define sets of networks with high and low network-level predictability by using the 2\% quantile-motivated thresholds (see Figure 2 in the main text).}\label{supFigure12}
	\end{center}
\end{figure*}

\begin{figure*}[!htbp]
	\begin{center}
		\subfloat[][]{
			\includegraphics[angle=-0,scale=0.25]{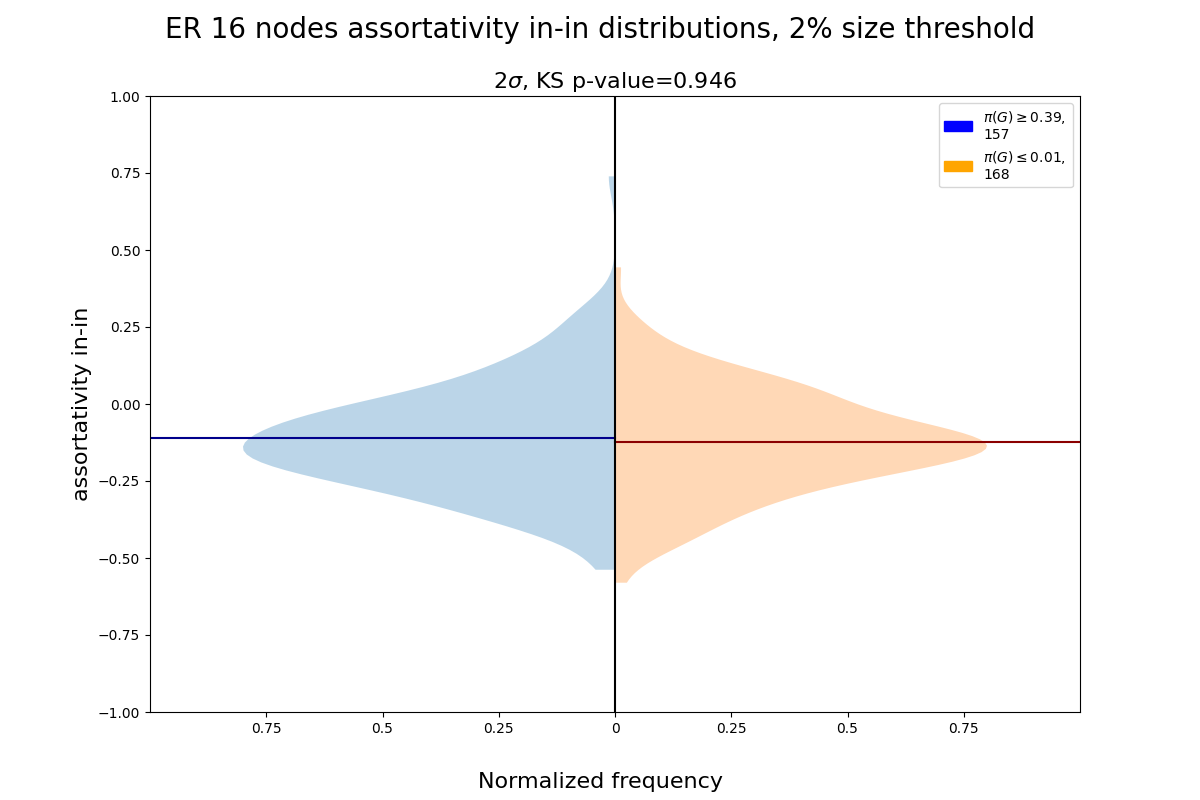}
		}
		\subfloat[][]{
			\includegraphics[angle=-0,scale=0.25]{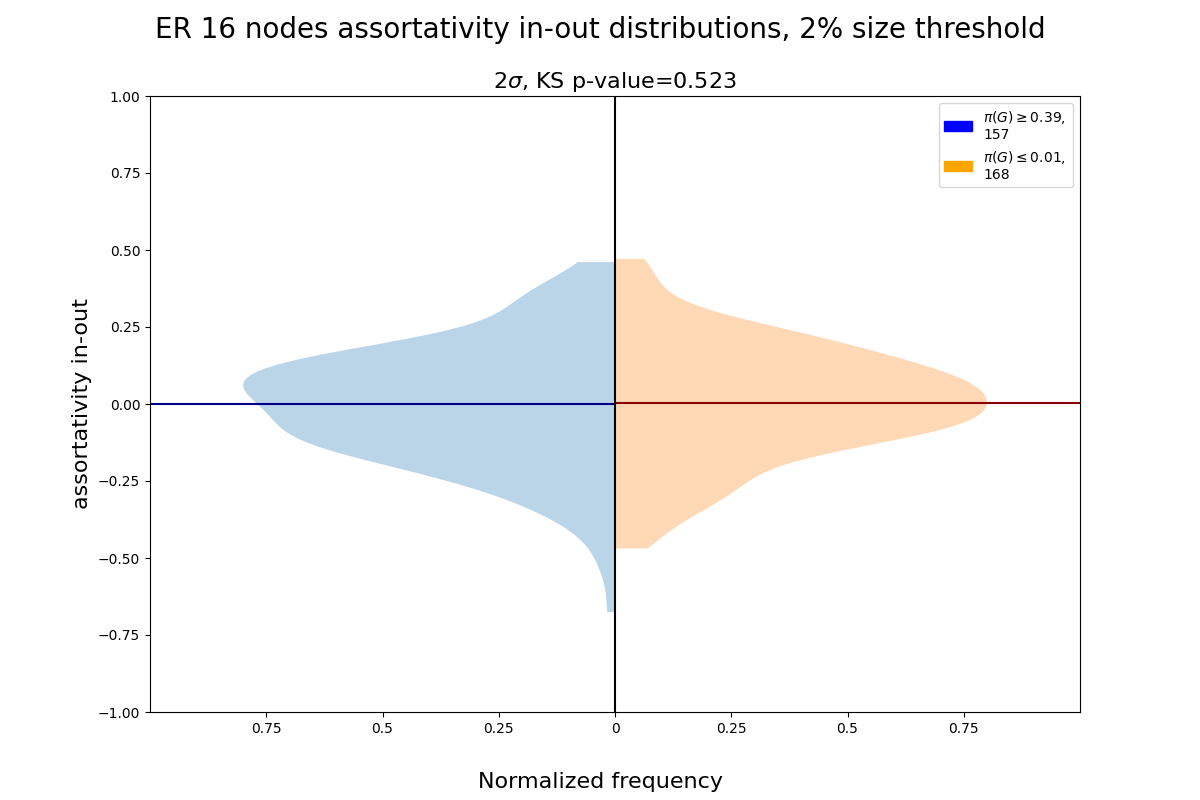}
		}\\
		\subfloat[][]{
			\includegraphics[angle=-0,scale=0.25]{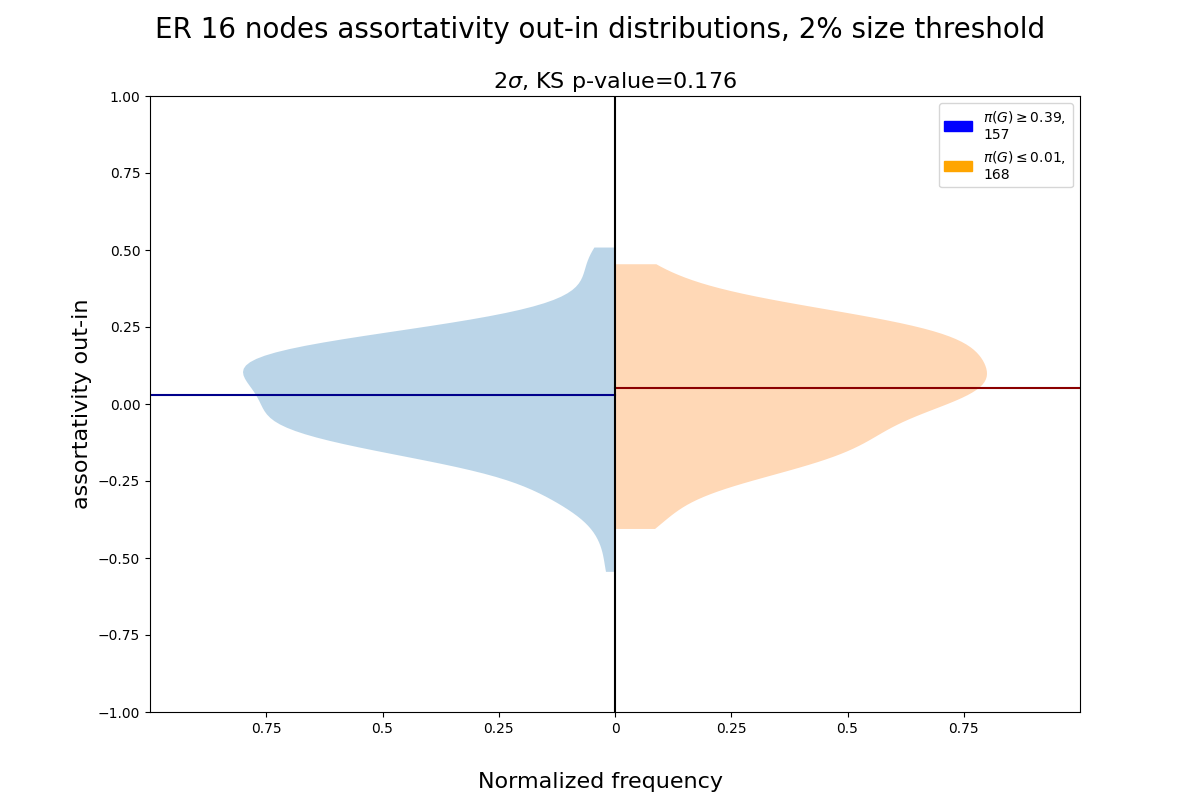}
		}
			\subfloat[][]{
		\includegraphics[angle=-0,scale=0.25]{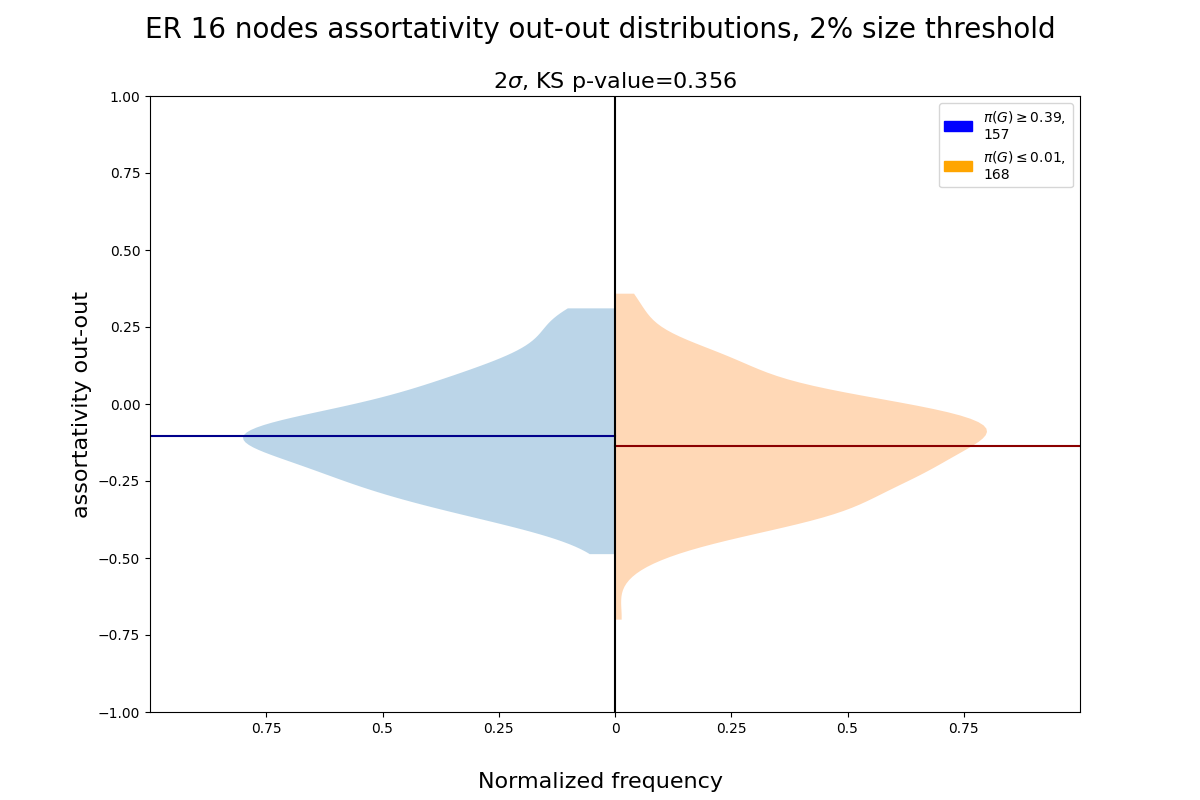}
	}
		\caption{Distributions of assortativity for all combinations of source-target node degree types for sets of networks with high and low network-level predictability. 10.000 random ER networks with 16 nodes, 32 positive and 32 negative links were considered. KS p-value corresponds to the p-value obtained from two-sample Kolmogorov–Smirnov test. We define sets of networks with high and low network-level predictability by using the 2\% quantile-motivated thresholds (see Figure 2 in the main text).} \label{supFigure13}
	\end{center}
\end{figure*}
\begin{figure*}[!hbtp]
	\begin{center}
		\includegraphics[angle=-0,scale=0.5]{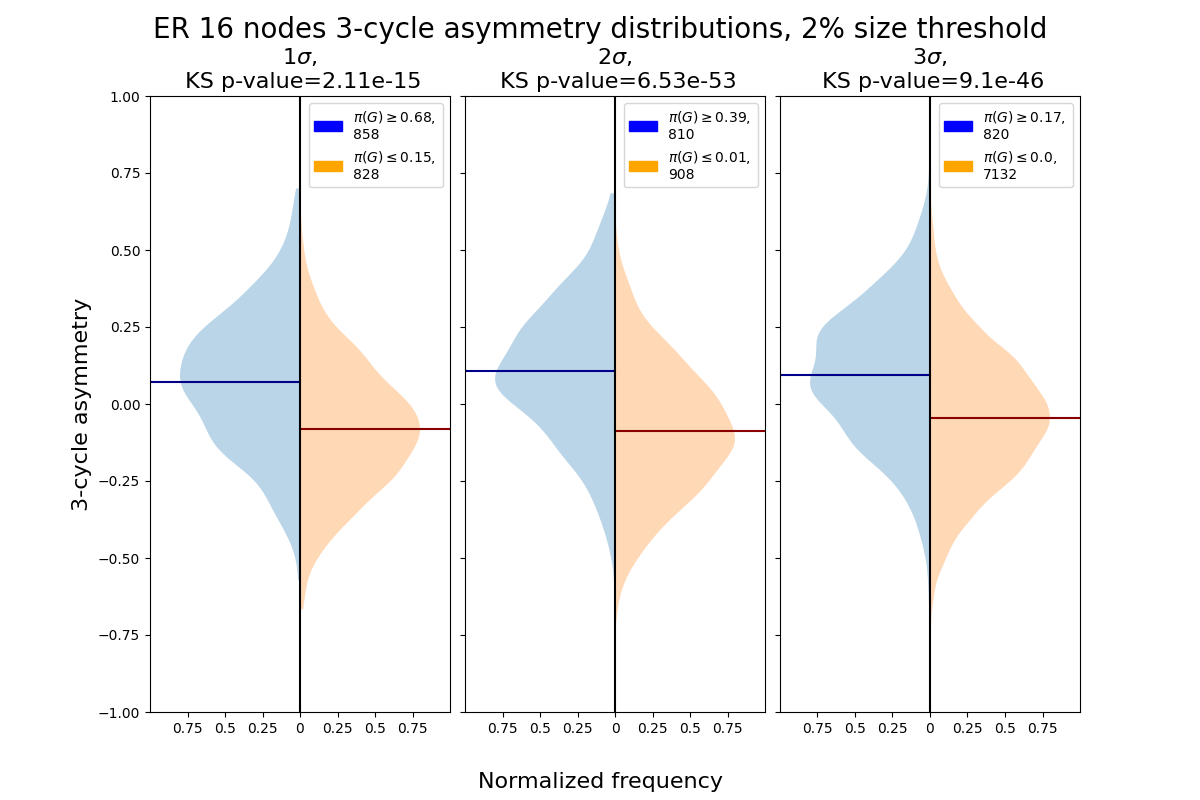}
		\caption{Distributions of 3-cycle asymmetry for sets of networks with high and low network-level predictability and exhaustive enumeration of initial conditions. A set of 50.000 ER networks on 16 nodes with 21 positive and 32 negative links has been considered. For each network dynamics was run starting from all 65.536 possible random initial states. Network-level predictability is displayed for thresholds of 1, 2 and 3 standard deviations away from randomness. KS p-value corresponds to the p-value obtained from two-sample Kolmogorov–Smirnov test. We define sets of networks with high and low network-level predictability by using the 2\% quantile-motivated thresholds (see Figure 2 in the main text).} \label{supFigure14}
	\end{center}
\end{figure*}
\begin{figure*}[!htbp]
	\begin{center}
		\subfloat[][]{
			\includegraphics[angle=-0,scale=0.25]{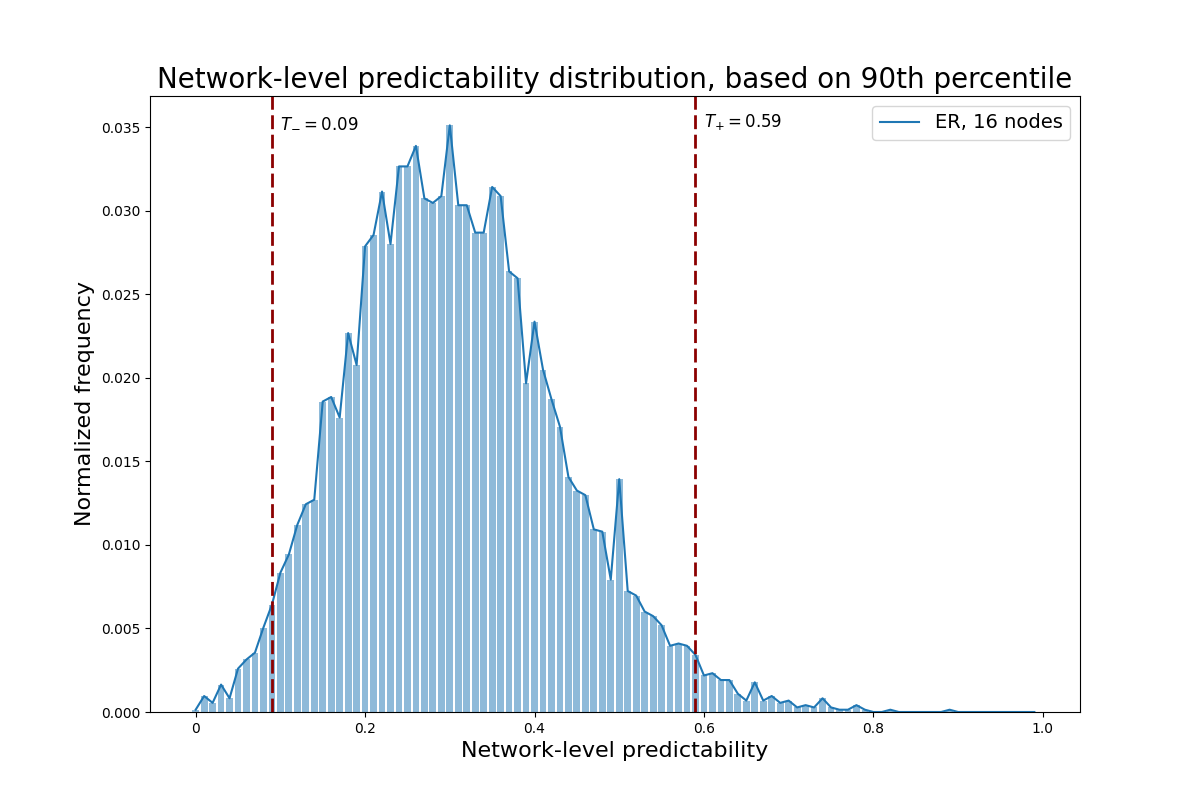}
		}
		\subfloat[][]{
			\includegraphics[angle=-0,scale=0.25]{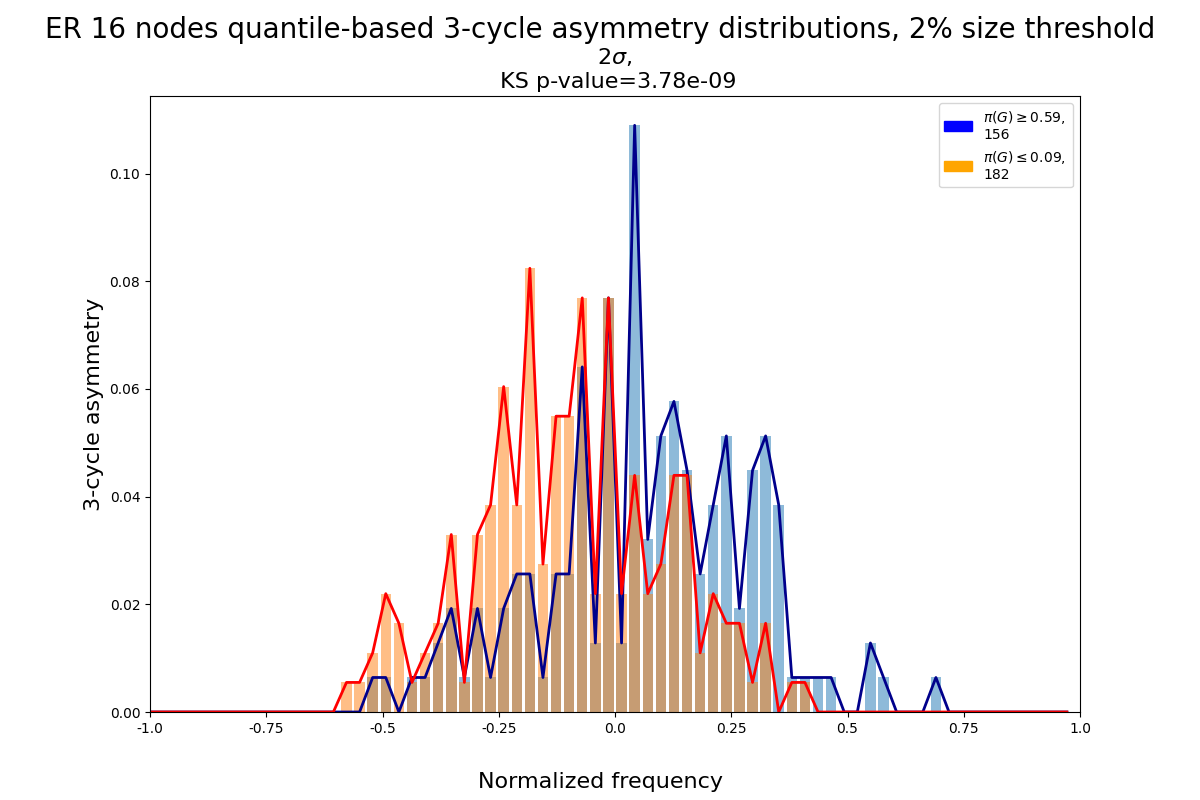}
		}
		\caption{(A) Distribution of network-level predictability defined based on 90th percentile, together with the thresholds $T_-=0.09$ and $T_+=0.59$ (motivated as in Figure 2C)for networks with high and low network-level predictability, respectively. Similarly to Figure 2A, attractor of a single network is said to be \textit{highly predictable} if it's predictability is at least in the 90-th percentile of the predictability distribution of shuffled attractors. 10.000 random ER networks with 16 nodes, 32 positive and 32 negative links were considered. (B) Histogram of 3-cycle asymmetry distributions for sets of networks with high (blue) and low (red) network-level predictability, defined using 90th percentile threshold(see above). KS p-value corresponds to the p-value obtained from two-sample Kolmogorov–Smirnov test } \label{supFigure15}
	\end{center}
\end{figure*}
\begin{figure*}[!hbtp]
	\begin{center}
		\includegraphics[angle=-0,scale=0.5]{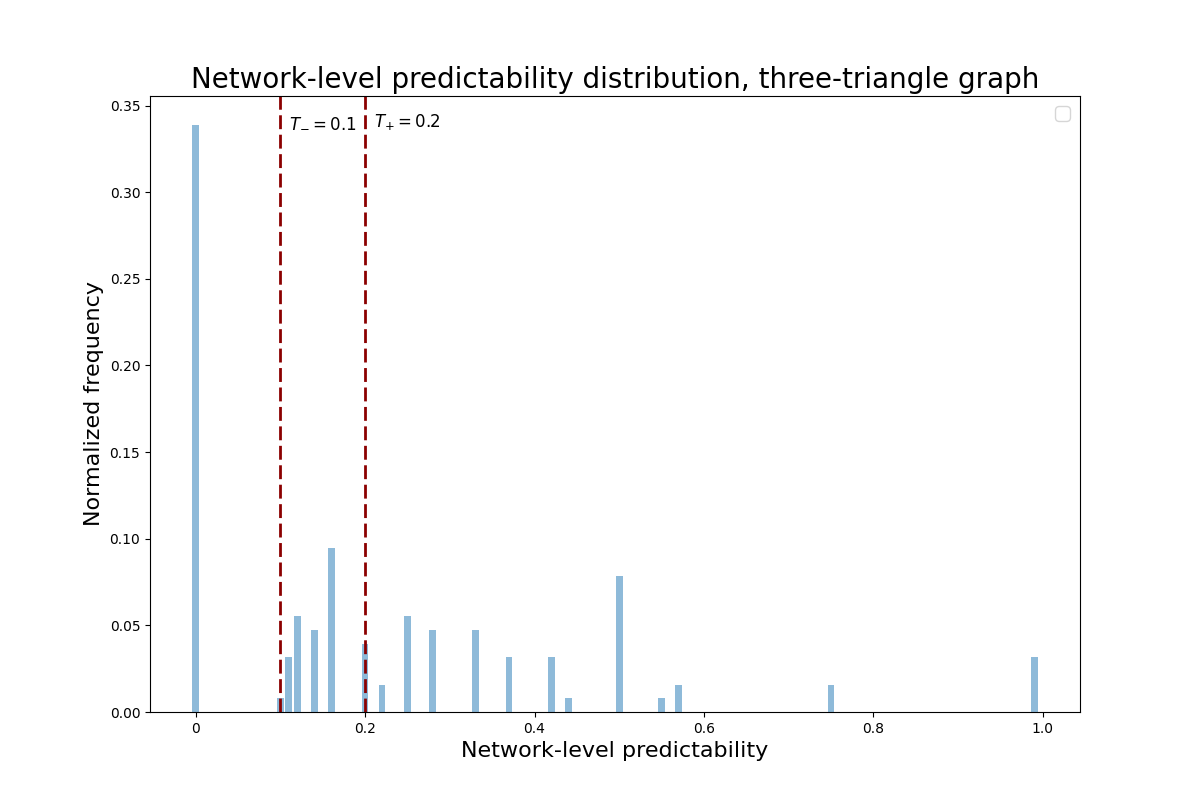}
		\caption{Distribution of network-level predictabilities for all 128 possible sign assignments on the directed topology in Figure 5A.} \label{supFigure16}
	\end{center}
\end{figure*}

\begin{center}
	\begin{table}[!hbtp]
		\caption{ Predictability $\pi(G)$, asymmetries $A_3(G)$ and $A_4(G)$ and agreement with theory for biological networks. Thresholds for high and low cycle asymmetry are $\pm0.25$. Low and high network-level predictability thresholds are $\pi(G) < 0.1$ and $\pi(G) > 0.3$.  Agreement is assessed as follows: if both $A_3(G)$ and $A_4(G)$ are below $-0.25$ for low or above $0.25$ for high network-level predictability, the results is $++$. If the previous condition holds for either $A_3(G)$ or $A_4(G)$, and the other asymmetry is between the thresholds, the result is a $+$. If network-level predictability and one of asymmetry values are extreme in the opposite sense, and the other asymmetry value is between the thresholds, the result is a $-$. Results of $-+$ and $--$ are defined similarly, and are not present in the table. \\}\label{table1}
		\resizebox{0.9\textwidth}{!}{
			\begin{tabular}{|l|c|c|c|c|c|c|c|}
				\hline
				\textbf{Model name}  & $\pi(G)$ & $A_3(G)$ & $A_4(G)$ & \textbf{3-cycle count} & \textbf{4-cycle count} & \textbf{Size} & \textbf{Agreement}\\
				\hline
Mammalian Cell Cycle 2006&0.01&0.20&-0.25&10&8&10&
\\ \hline Budding Yeast Cell Cycle 2009&0.07&-0.33&-0.76&9&25&18&++
\\ \hline CD4+ T Cell Differentiation and Plasticity&0.07&0.33&0.04&51&150&18&-
\\ \hline Oxidative Stress Pathway&0.08&-1.00&-1.00&2&1&19&++
\\ \hline Budding Yeast Cell Cycle&0.11&1.00&-0.27&3&11&20&
\\ \hline Arabidopsis thaliana Cell Cycle&0.15&-0.52&0.00&21&54&14&
\\ \hline emt26network&0.16&0.80&0.93&10&56&26&
\\ \hline Lac Operon&0.21&nan&1.00&0&2&13&
\\ \hline T-LGL Survival Network 2011 Reduced Network&0.31&-0.33&0.00&3&6&18&-
\\ \hline B cell differentiation&0.32&1.00&1.00&6&4&22&++
\\ \hline gonadalsexdet&0.34&0.64&0.62&28&78&19&++
\\ \hline Tumour Cell Invasion and Migration&0.36&0.53&0.27&64&222&32&++
\\ \hline sclcnetwork&0.78&0.50&0.37&482&3745&33&++
\\ \hline gastricnetwork&0.79&0.42&0.30&38&88&48&++
\\ \hline Aurora Kinase A in Neuroblastoma&0.95&0.20&0.43&5&7&23&+
\\ \hline 
		\end{tabular}}	
	\end{table}
\end{center}
	\clearpage
	\bibliography{supplements}